\documentclass[journal=jctcce,manuscript=article,layout=twocolumn]{achemso}

\usepackage[version=4]{mhchem}
\usepackage[T1]{fontenc}
\usepackage{amsmath}
\usepackage{amssymb}
\usepackage{bm}
\usepackage{graphicx}
\usepackage{booktabs}
\usepackage{xcolor}
\usepackage{siunitx}
\usepackage{hyperref}

\newcommand{\Ch}{\mathbf{C}_h}
\newcommand{\T}{\mathbf{T}}
\newcommand{\av}{\mathbf{a}_1}
\newcommand{\bv}{\mathbf{a}_2}
\newcommand{\ntbuilder}{\textsf{NTBuilder}}

\author{Marcelo Lopes Pereira Junior}
\email{marcelo.lopes@unb.br}
\affiliation{Department of Electrical Engineering, College of Technology, University of Bras\'ilia, 70910-900, Bras\'ilia, Federal District, Brazil}

\title{NTBuilder: Commensurate Construction of Nanotubes from Arbitrary Two-Dimensional Crystals and a Catalog of 20 Million Structures}


\begin{document}

\begin{tocentry}
\centering
\includegraphics[width=0.9\linewidth]{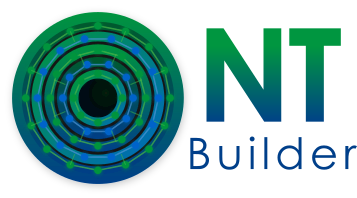}
\end{tocentry}

\begin{abstract}
Although tens of thousands of two-dimensional (2D) crystals are now known, the nanotubes they can form are usually constructed for individual materials, using procedures derived for graphene. Here, we develop a general theory of the rolling construction for arbitrary 2D crystals and implement it in \ntbuilder{} (NanoTube Builder), an open-source library with desktop and web interfaces. We express the size of the exact unit cell of a tube through the reduced form of a single rational number built from the lattice metric, which explains why triangular and square lattices always close with small cells, whereas rectangular, centered rectangular, and oblique lattices can require cells larger by many orders of magnitude. The optimal approximate cells are the best rational approximations of a second number, so that a cell small enough for first-principles calculations can be chosen with a controlled departure from periodicity, as in the biphenylene network, where such a cell is eight orders of magnitude smaller than the exact one. The 17 plane groups reduce to seven distinct chirality maps once rolling is taken into account, the rolling sense of layers without a horizontal mirror defines distinct tubes, and bonds formed or broken by curvature are detected pair by pair against the flat layer. The same framework assembles multiwalled tubes whose walls share one axial period, periodic bundles, and strained or twisted structures that remain periodic. Applied to 46,403 distinct 2D systems, it produced a public catalog of 20,334,372 nanotubes, of which 57.1\% show no bond change upon rolling. All structures can be exported directly as input for first-principles and molecular dynamics codes.
\end{abstract}

\section{Introduction}
\label{sec:intro}

Two-dimensional (2D) crystals have become one of the most productive platforms of materials science since the isolation of graphene,\cite{Novoselov2004} whose massless charge carriers,\cite{Novoselov2005, Zhang2005} mechanical strength,\cite{Lee2008} and atomic thickness revealed how strongly dimensionality reshapes the properties of a solid.\cite{Geim2007, CastroNeto2009} The family has since grown to include hexagonal boron nitride (h-BN), transition metal dichalcogenides, MXenes, and a rapidly expanding set of carbon allotropes such as the biphenylene network, Irida-graphene, and penta-graphene.\cite{Bhimanapati2015, Ajayan2016, Manzeli2017, Fan2021, PereiraJunior2023, Zhang2015} In parallel, high-throughput exfoliation and screening studies have organized candidate monolayers into public databases that together list more than ten thousand entries,\cite{Mounet2018, Campi2023, Haastrup2018, Gjerding2021, Zhou2019, Choudhary2017, Choudhary2020} and a random search guided by group and graph theory has identified 1,114 previously unreported planar \textit{sp}$^2$ carbon allotropes.\cite{Shi2021} The number of 2D crystals available for study already far exceeds what can be examined individually.

Rolling a monolayer into a seamless cylinder is the most direct way to obtain one-dimensional structures from these layers. Carbon nanotubes were identified by Iijima in 1991,\cite{Iijima1991} and the chiral-vector construction introduced by Hamada \textit{et al.} and Saito \textit{et al.} showed that a pair of integers, the chiral indices $(n,m)$, fixes the diameter, the helicity, and the electronic character of each tube.\cite{Hamada1992, Saito1992, White1993, Saito1998, Charlier2007} Inorganic nanotubes followed with the synthesis of \ce{WS2} tubes by Tenne \textit{et al.},\cite{Tenne1992} and tubes of BN, \ce{MoS2}, and many other layered compounds have since been produced or predicted.\cite{Chopra1995, Seifert2000, Nath2001, Tenne2004} Beyond single tubes, nanotubes occur as multiwalled structures and as close-packed bundles,\cite{Iijima1991, Thess1996} Janus monolayers such as \ce{MoSSe}\cite{Lu2017, Zhang2017} introduce a rolling sense that determines which face lies inside the tube,\cite{Luo2019, Evarestov2020} and sheets that do not close form nanoscrolls.\cite{Braga2004, Perim2013}

For graphene, the construction reduces to closed-form expressions in the chiral indices, which give the translation vector and the number of atoms in the unit cell, and the symmetry of every tube is known in full.\cite{White1993, Damnjanovic1999, Saito1998} These expressions rely on two properties that a general 2D crystal does not share. The first is commensurability, since a translation vector exactly perpendicular to the chiral vector exists only when a ratio built from the lattice metric is rational, and the size of the resulting cell is then set by the denominator of that ratio. The second is symmetry, since the set of distinct tubes depends on the point group of the decorated layer and on which of its two faces is turned inward, and not only on the Bravais lattice. A third factor arises when the layer has a finite thickness, because rigid rolling compresses the inner face and stretches the outer one, which can create or break bonds that do not exist in the flat sheet.

General-purpose crystallographic libraries such as pymatgen and the Atomic Simulation Environment provide the data structures to represent periodic systems,\cite{Ong2013, Larsen2017} and visualization programs such as VESTA and VMD are routinely used to inspect them.\cite{Momma2011, Humphrey1996} Among the dedicated generators, TubeGen rolls honeycomb sheets and replicates the tubes into periodic bundles,\cite{Frey2011} CoNTub joins carbon nanotubes of different chiralities,\cite{Melchor2004} Atomsk builds and transforms atomic systems for simulation codes,\cite{Hirel2015} and the generator of the CRYSTAL code exploits the helical symmetry of single- and multiwalled tubes of layered compounds.\cite{Noel2010, Marana2021} Of the recent tools that start from an arbitrary 2D cell, Chiraltube accepts an approximate translation vector within a tolerance on its integer indices and rolls multilayered sheets by expanding the outer and compressing the inner atomic layers,\cite{DeAlbornozCaratozzolo2024} and NanoTube Construct replicates the sheet until the mismatch falls below a threshold and relaxes the result with force fields.\cite{Kolokathis2024} Line-group theory gives the full symmetry of carbon nanotubes\cite{Damnjanovic1999} and the periodicity condition of tubes rolled from rectangular\cite{Milosevic2006} and arbitrary lattices.\cite{Damnjanovic2007} None of these tools enumerates the complete trade-off between the size of the cell and its periodicity, reduces the chirality map by the symmetry of the decorated layer including the exchange of its faces, detects the bonds formed or broken by curvature, or builds walls and bundles that share one exact axial period. Studies of nanotubes derived from non-hexagonal or low-symmetry monolayers therefore rely on scripts written for a single material, and a systematic account of the nanotubes that the known 2D crystals can form has not been available.

\begin{figure*}[!t]
  \centering
  \includegraphics[width=0.8\linewidth]{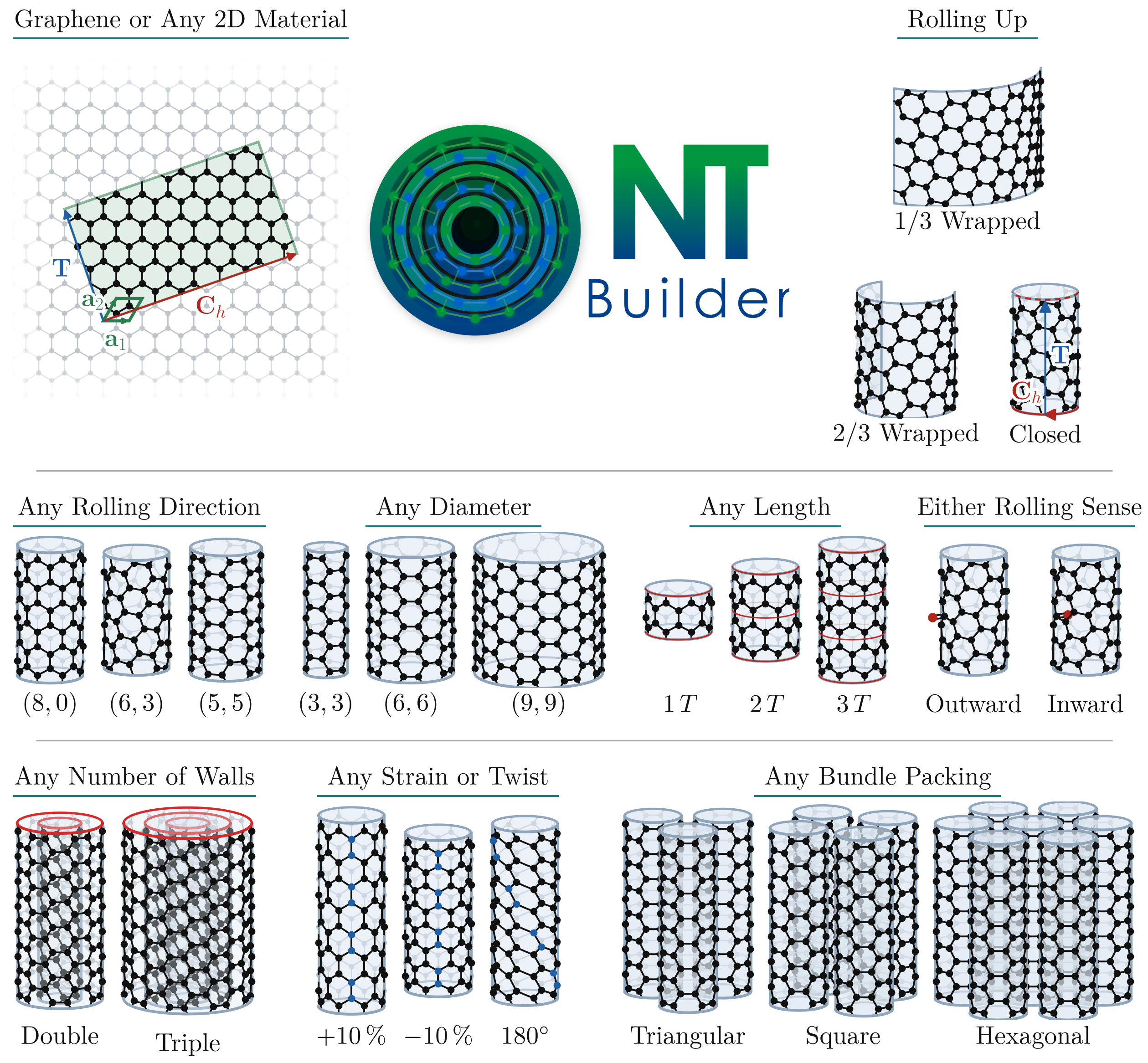}
  \caption{Overview of the constructions available in \ntbuilder{}, illustrated on graphene, with the flat sheet, two intermediate wraps, and the closed tube (top), tubes of three chiral angles (middle), and multiwalled tubes, bundle packings, axial strain, and torsion (bottom). Wall and bundle spacings are compressed for display.}
  \label{fig:roll}
\end{figure*}

In this work, we present \ntbuilder{} (NanoTube Builder), an open-source library with desktop and web interfaces that builds nanotubes from any 2D crystal, together with the theory on which it rests. We write the size of the exact unit cell of a tube as a cost law set by the reduced form of a single rational number, show that every approximate cell lies on a trade-off front given by the best rational approximations of a second number, show that the 17 plane groups induce seven distinct chirality maps once rolling is taken into account, and detect the bonds formed or broken by curvature pair by pair against the flat layer. The same framework assembles multiwalled tubes and bundles that share one axial period, applies axial strain and closing twists, and was used to build a catalog of 20,334,372 nanotubes from 46,403 distinct 2D systems. Figure~\ref{fig:roll} gives an overview of the constructions available in \ntbuilder{}, illustrated on graphene, from the rolling of the flat sheet and the choice of the chiral angle to multiwalled tubes, bundles, axial strain, and torsion.

\section{Rolling a 2D Crystal}
\label{sec:rolling}

Rolling a 2D crystal into a tube raises three questions, which the following subsections address in the order in which they arise. The first concerns the periodicity of the tube along its axis and the size of the cell that achieves it, the second concerns the number of distinct tubes that the symmetry of the layer allows, and the third concerns the bonds that curvature creates or breaks.

\subsection{Periodicity and Its Cost}
\label{sec:cost}

A nanotube is defined by the lattice vector of the sheet that becomes its circumference. For a 2D Bravais lattice with primitive vectors $\av$ and $\bv$, the chiral vector is $\Ch = n\av + m\bv$ with $(n,m)\in\mathbb{Z}^2$, and the metric enters through the three numbers $A = \av\cdot\av$, $B = \av\cdot\bv$, and $C = \bv\cdot\bv$. The diameter of the tube is
\begin{equation}
  D = \frac{|\Ch|}{\pi} = \frac{1}{\pi}\sqrt{n^2A + 2nmB + m^2C},
  \label{eq:diameter}
\end{equation}
\noindent which depends on the lattice and not on the atoms it carries. Each atom of the layer is described by its in-plane coordinates $\xi$ along $\Ch$ and $\eta$ perpendicular to it, together with its height $\zeta$ above the mid-plane of the layer. Rolling maps these coordinates onto a cylinder of radius $R = |\Ch|/2\pi$,
\begin{equation}
  \mathbf{r} = \Big( (R + \sigma\zeta)\cos\frac{2\pi\xi}{|\Ch|},\;
                     (R + \sigma\zeta)\sin\frac{2\pi\xi}{|\Ch|},\; \eta \Big),
  \label{eq:rolling}
\end{equation}
\noindent where $\sigma = \pm 1$ selects which face of the layer points outward. Eq.~\ref{eq:rolling} preserves every length along the axis and rescales the circumferential arc lengths at height $\zeta$ by $(R + \sigma\zeta)/R$.

\begin{figure*}[!t]
  \centering
  \includegraphics[width=0.8\linewidth]{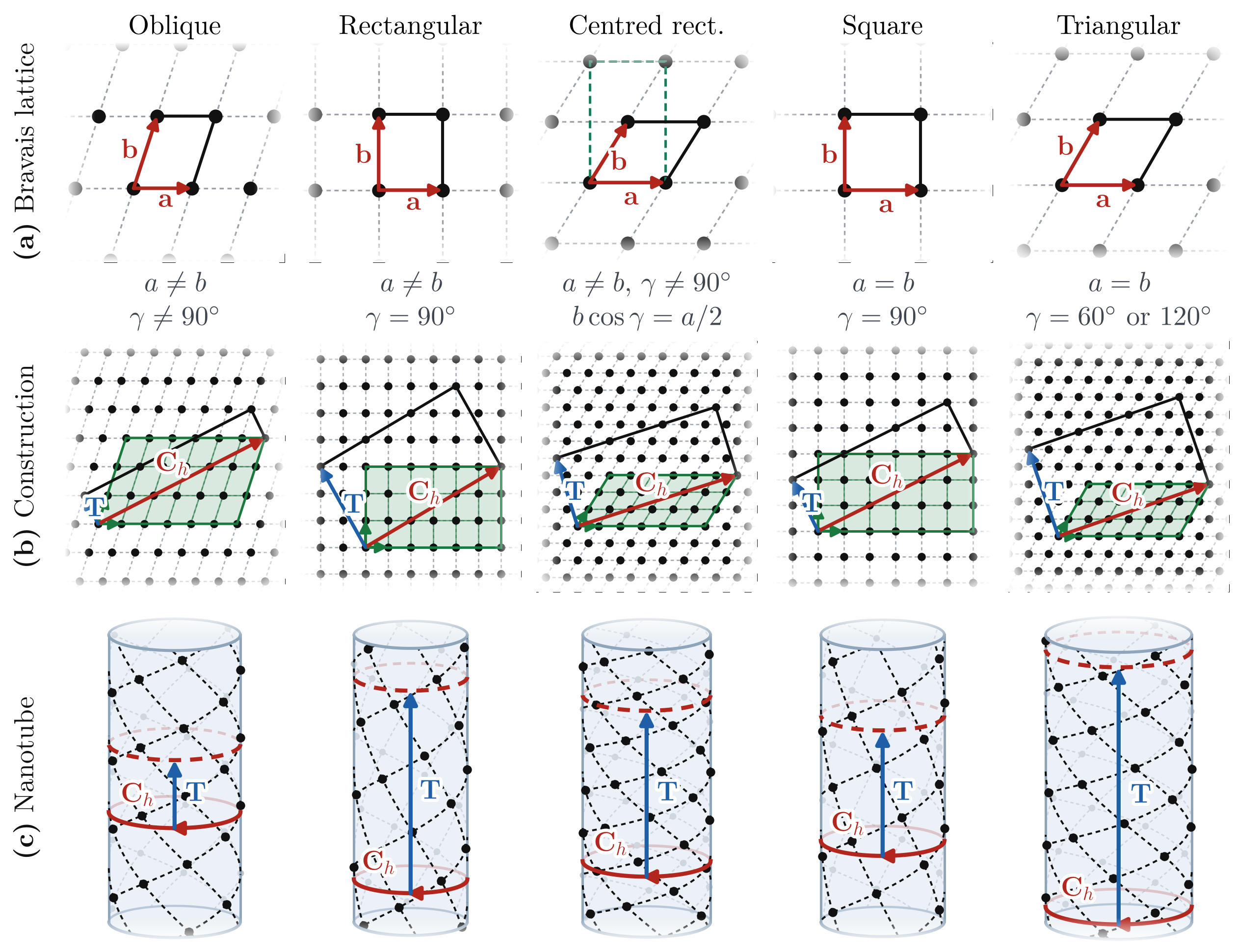}
  \caption{The five 2D Bravais lattices (a), the chiral vector $\Ch$, with the same indices on every lattice, and the translation vector $\T$ (b), and the corresponding tubes (c).}
  \label{fig:bravais}
\end{figure*}

Eq.~\ref{eq:diameter} already separates two questions, since the geometry of a tube, namely its diameter, its axial period, and the size of its unit cell, is determined by the metric $(A, B, C)$ alone, so that all 2D crystals sharing a Bravais lattice share these quantities and differ only in the number of atoms per cell, whereas the number of distinct tubes a crystal can form is determined by the symmetry of the decorated layer, which is the subject of Section~\ref{sec:symmetry}. Figure~\ref{fig:bravais} illustrates the first point by building a tube with the same chiral indices on one representative of each of the five 2D Bravais lattices.

The same indices give tubes of different periodicity on the five lattices. On the square and triangular lattices, a translation vector exactly perpendicular to the chiral vector always exists, and the number of primitive cells in the unit cell does not depend on the lattice constants. On the rectangular, centered rectangular, and oblique lattices, a small exact cell exists only for particular ratios of the lattice parameters, and a bounded search in general returns a cell that is only approximately periodic. This contrast follows from the arithmetic of the metric, since the translation vector $\T = t_1\av + t_2\bv$ must satisfy $\Ch\cdot\T = 0$, which reads
\begin{equation}
  t_1\,(nA + mB) + t_2\,(nB + mC) = 0,
  \label{eq:orthogonality}
\end{equation}
\noindent so an integer solution exists if and only if the ratio
\begin{equation}
  r = \frac{nB + mC}{nA + mB}
  \label{eq:ratio}
\end{equation}
\noindent is rational. Writing $r = p/q$ in lowest terms, the shortest solution is $(t_1, t_2) = (-p, q)$, and the number of primitive cells in the rectangle spanned by $\Ch$ and $\T$ is the ratio of the two areas,
\begin{equation}
  N = \frac{|\Ch \times \T|}{|\av \times \bv|} = |n t_2 - m t_1|
    = |nq + mp|.
  \label{eq:costlaw}
\end{equation}
\noindent Eq.~\ref{eq:costlaw} is the cost law of an exact tube, a compact form of the periodicity condition that line-group theory derives for rectangular\cite{Milosevic2006} and arbitrary\cite{Damnjanovic2007} lattices. The size of the exact unit cell is governed by the reduced ratio, through its denominator $q$ and numerator $p$, and not by the Bravais class as such.

\begin{figure*}[!t]
  \centering
  \includegraphics[width=0.8\linewidth]{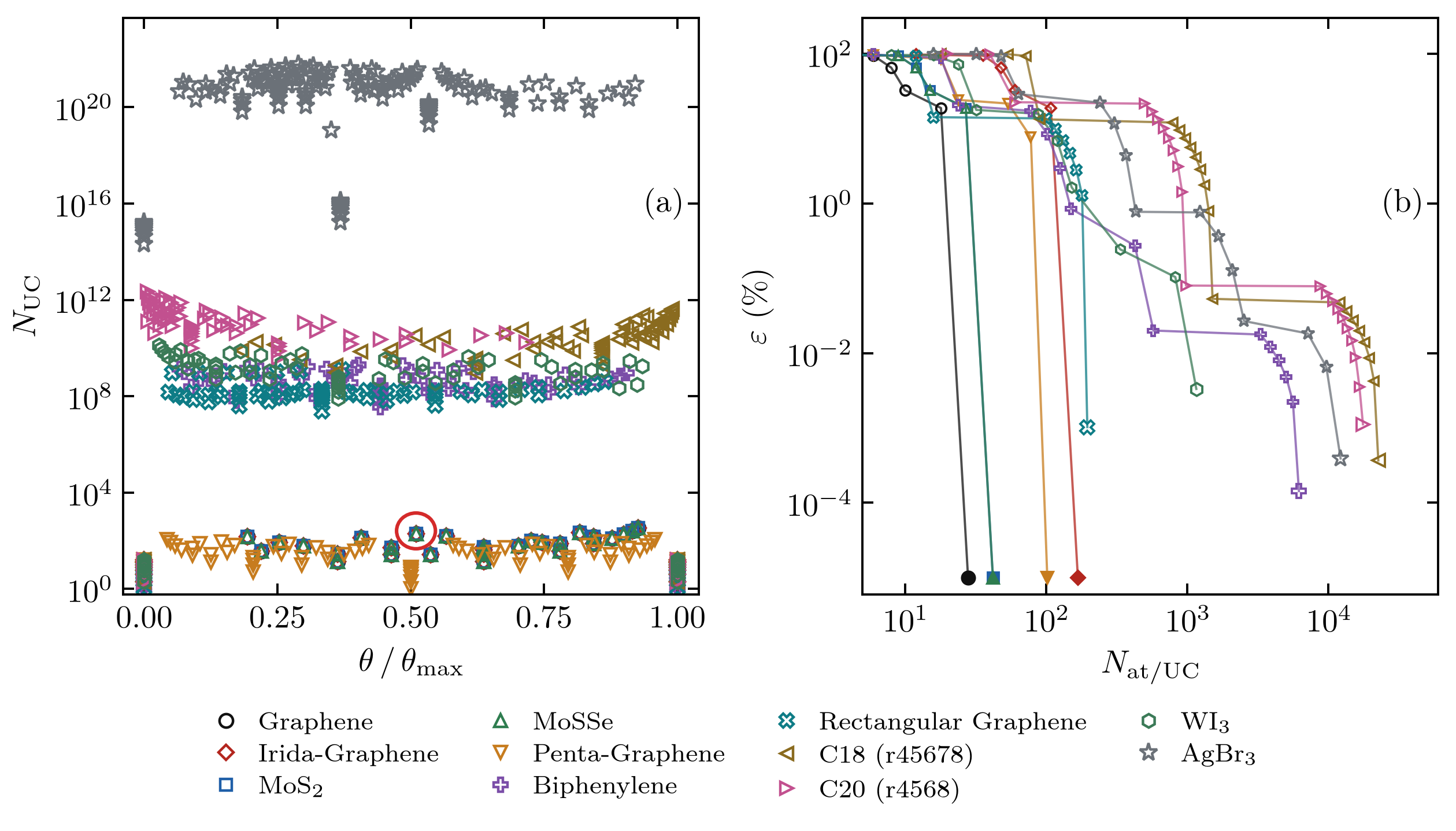}
  \caption{Number of primitive cells $N_{\mathrm{UC}}$ in the exact unit cell as a function of the normalized chiral angle $\theta/\theta_{\max}$ (a) and periodicity residual $\varepsilon$ as a function of the number of atoms per cell for the $(4,1)$ tube (b), for the systems of Table~\ref{tab:systems}.}
  \label{fig:exactness}
\end{figure*}

The classical results are special cases of Eq.~\ref{eq:costlaw}. On the triangular lattice, $A = C = 2B$ gives $r = (n+2m)/(2n+m)$ and
\begin{equation}
  N_{\mathrm{tri}} = \frac{2\,(n^2 + nm + m^2)}{\gcd(2n+m,\, n+2m)},
  \label{eq:triangular}
\end{equation}
\noindent which, multiplied by the two atoms of the graphene basis, reproduces the textbook count $4(n^2+nm+m^2)/d_R$ with $d_R = \gcd(2n+m, 2m+n)$.\cite{Saito1998} On the square lattice, $r = m/n$ and $N_{\mathrm{sq}} = (n^2+m^2)/\gcd(n,m)$. In both cases $r$ is rational for every pair of indices and independent of the lattice constant, so these two classes always admit exact cells, whose sizes are common to all their members. On a rectangular lattice with $C/A = u/v$ in lowest terms, $r = mu/(nv)$ and
\begin{equation}
  N_{\mathrm{rect}} = \frac{n^2 v + m^2 u}{\gcd(mu,\, nv)},
  \label{eq:rectangular}
\end{equation}
\noindent so the cell size depends on the lattice parameters themselves, and the same holds for the centered rectangular and oblique lattices through $\cos\gamma$. Since every number stored in floating point is rational, the rationality of $r$ cannot be decided numerically, and what a program can report is whether an exact $\T$ was found within a given search bound. Figure~\ref{fig:exactness}(a) applies Eq.~\ref{eq:costlaw} to every irreducible index pair with $n, m \le 8$ for the eleven systems of Table~\ref{tab:systems}, which cover the five Bravais classes.

\begin{table*}[!h]
  \centering
  \caption{Lattice parameters, number of atoms in the primitive cell $n_{\mathrm{at}}$, Bravais lattice, order $|G|$ of the group acting on the chiral indices, and width of the irreducible sector of the chirality map for the systems of Figure~\ref{fig:exactness}.}
  \label{tab:systems}
  \begin{tabular}{lrrrrlrr}
    \toprule
    System & $a$ (\AA) & $b$ (\AA) & $\gamma$ ($^\circ$) & $n_{\mathrm{at}}$
      & Lattice & $|G|$ & Sector ($^\circ$) \\
    \midrule
    Graphene              &  2.460 &  2.460 &  60.00 &  2 & triangular            & 12 &  30 \\
    Irida-graphene        &  6.340 &  6.340 &  60.00 & 12 & triangular            & 12 &  30 \\
    \ce{MoS2}             &  3.192 &  3.192 &  60.00 &  3 & triangular            & 12 &  30 \\
    \ce{MoSSe}            &  3.253 &  3.253 &  60.00 &  3 & triangular            & 12 &  30 \\
    Penta-graphene        &  3.630 &  3.630 &  90.00 &  6 & square                &  4 &  90 \\
    Biphenylene network   &  4.519 &  3.771 &  90.00 &  6 & rectangular           &  4 &  90 \\
    Rectangular graphene  &  4.261 &  2.460 &  90.00 &  4 & rectangular           &  4 &  90 \\
    C18 (r45678)          & 10.781 & 10.781 &  25.38 & 18 & centered rectangular  &  4 &  90 \\
    C20 (r4568)           & 14.212 & 14.212 & 164.13 & 20 & centered rectangular  &  4 &  90 \\
    \ce{WI3}              &  6.523 &  6.523 & 114.52 &  8 & centered rectangular  &  4 &  90 \\
    \ce{AgBr3}            & 16.608 &  8.318 & 113.76 & 16 & oblique               &  2 & 180 \\
    \bottomrule
  \end{tabular}
\end{table*}

The four triangular systems of Figure~\ref{fig:exactness}(a), graphene, Irida-graphene, \ce{MoS2}, and \ce{MoSSe}, fall on a single set of points, with $N_{\mathrm{UC}}$ between 2 and 338 and a median of 33 cells, because Eq.~\ref{eq:triangular} contains neither the lattice constant nor the basis. Penta-graphene, the only square system, closes with at most 113 cells. In contrast, the biphenylene network and the rectangular cell of graphene reach medians of about $2.8\times10^{8}$ and $1.1\times10^{8}$ cells, and the centered rectangular and oblique systems reach medians from $10^{9}$ cells for the \ce{WI3} monolayer to $10^{21}$ cells for \ce{AgBr3}, both taken from C2DB,\cite{Haastrup2018, Gjerding2021} set by the lattice constants and angles at the precision used here, four decimals in the lengths. C18 and C20, the two carbon allotropes with the smallest and largest lattice angles in the database of Shi \textit{et al.},\cite{Shi2021} belong to this last group, and their angles of 25.38 and $164.13^\circ$ lead to the same $90^\circ$ sector because both lattices are centered rectangular. Rectangular graphene illustrates the role of these decimals, since its ideal ratio $C/A = 1/3$ is rational and gives small exact cells through Eq.~\ref{eq:rectangular}, whereas the stored lattice constants of 4.261 and 2.460~\AA\ turn it into 0.3333091, so that the exact cell of the stored structure is set by the truncated input. The same holds for the biphenylene network, whose lattice constants of 4.519 and 3.771~\AA\ are stored with three decimals. In the classes with mirrors, each sector ends on mirror lines, where $\T$ is a short lattice vector and the cell reduces to a few primitive cells.

When the exact cell is too large for practical use, periodicity must be traded for cell size. We quantify the departure from periodicity by the residual
\begin{equation}
  \varepsilon = \frac{|\Ch\cdot\T|}{|\Ch|\,|\T|},
  \label{eq:residual}
\end{equation}
\noindent the cosine of the angle between $\Ch$ and $\T$. \ntbuilder{} places the atoms of an approximate cell at their fractional coordinates in the basis $(\Ch, \T)$, which amounts to a simple shear of the strip by the angle $\arcsin\varepsilon$ that makes $\T$ perpendicular to $\Ch$ while keeping both lengths, and changes bond lengths by at most about $\varepsilon/2$. The relevant cells are those for which no smaller cell has a lower residual. Writing $\Ch = \nu\,\Ch'$ with $\nu = \gcd(n,m)$ and $\Ch' = n'\av + m'\bv$ primitive, and choosing any lattice vector $\T_1$ with $n' t_2 - m' t_1 = 1$, every cell of $\nu\,c$ primitive cells has a translation vector of the form $c\,\T_1 - k\,\Ch'$, up to sign, with integer $k$. All such vectors share the component $v = c\,v_1$ perpendicular to $\Ch$, with $v_1 = S/|\Ch'|$ and $S = |\av \times \bv| = \sqrt{AC - B^2}$ the area of the primitive cell, while their component along $\Ch$ is
\begin{equation}
  u = |\Ch'|\,(c\,\alpha - k),
  \label{eq:front}
\end{equation}
\noindent where
\begin{equation}
  \alpha = \frac{\T_1\cdot\Ch'}{|\Ch'|^{2}},
  \label{eq:alpha}
\end{equation}
\noindent and the residual becomes $\varepsilon = x/\sqrt{1+x^2}$ with $x = |u|/v$, a strictly increasing function of $|\alpha - k/c|$. The trade-off front between size and periodicity is therefore the sequence of best rational approximations of the first kind of $\alpha$,\cite{Khinchin1964} and the residual decreases only at the denominators of the front. The convergents of $\alpha$, which belong to the front except for the first one, $\lfloor\alpha\rfloor/1$, when the fractional part of $\alpha$ exceeds $1/2$, also bound how fast the residual decays. For every convergent $k/c$, $|c\,\alpha - k| < 1/c$,\cite{Khinchin1964} so that $x < |\Ch'|^2/(c^2 S)$, and eliminating $c$ through $|\T|^2 = v^2(1 + x^2)$ yields
\begin{equation}
  \varepsilon\sqrt{1 - \varepsilon^2} < \frac{S}{|\T|^2}.
  \label{eq:bound}
\end{equation}
\noindent Since a cell of $N = \nu\,c$ primitive cells satisfies $|\Ch \times \T| = N S$, Eq.~\ref{eq:bound} is equivalent to $\varepsilon < \pi^2 D^2 \sqrt{1-\varepsilon^2}/(S N^2)$, and in particular $\varepsilon < \pi^2 D^2/(S N^2)$. The residual of the convergent cells therefore decays at least as $S/|\T|^2$ with the period and as $\pi^2 D^2/(S N^2)$ with the number of primitive cells, for any lattice and any basis, and for almost every $\alpha$ this exponent cannot be improved.\cite{Khinchin1964} The residual vanishes at the reduced denominator of $\alpha$, where $\nu\,c$ equals the $N$ of Eq.~\ref{eq:costlaw}, and a search bounded by $|t_2| \le t_{\max}$ in Eq.~\ref{eq:orthogonality} reaches that cell only when $t_{\max}$ reaches the denominator $q$ of $r$. For the $(4,1)$ tube of the biphenylene network this denominator is 81,685,444, far beyond the range of Figure~\ref{fig:exactness}(b), where the triangular and square systems close within a few cells and all the others keep a finite residual. \ntbuilder{} enumerates the front lazily in increasing cell size, so that a user can request the smallest cell within a residual tolerance or the most periodic cell below a given length without computing the cells beyond the answer.

\begin{figure*}[!t]
  \centering
  \includegraphics[width=\linewidth]{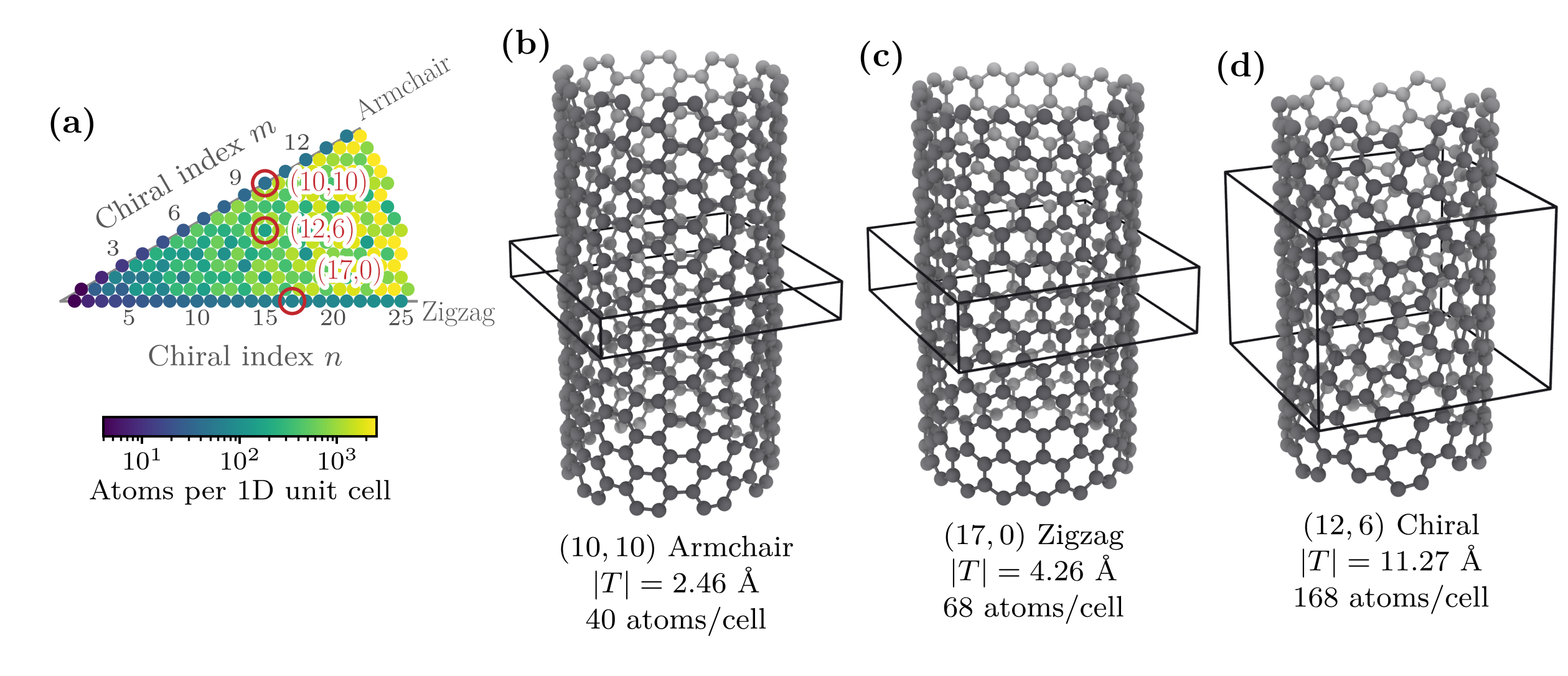}
  \caption{Chirality map of graphene up to $D = 20$~\AA, colored by the number of atoms per unit cell (a), and the $(10,10)$, $(17,0)$, and $(12,6)$ tubes with one unit cell highlighted (b--d).}
  \label{fig:graphene}
\end{figure*}

Graphene provides a check of this construction, since all its tubes close exactly and their periods are known in closed form. Figure~\ref{fig:graphene} shows the irreducible $30^\circ$ sector of its chirality map up to $D = 20$~\AA, with each of its 218 index pairs colored by the number of atoms in the unit cell, together with three tubes of different chiral angles.

The armchair $(10,10)$ and zigzag $(17,0)$ tubes close with periods of 2.46 and 4.26~\AA\ and cells of 40 and 68 atoms, while the chiral $(12,6)$ tube needs 168 atoms in 11.27~\AA. These values coincide with Eq.~\ref{eq:triangular}, which gives 20, 34, and 84 primitive cells, and with the periods $|\T| = a$ and $\sqrt{3}\,a$ of the armchair and zigzag families.\cite{Saito1998} The coloring of the map follows Eq.~\ref{eq:triangular}, with the smallest cells along the two achiral edges of the sector and the largest ones inside it, where $\gcd(2n+m, n+2m)$ is small.

\begin{figure*}[!t]
  \centering
  \includegraphics[width=\linewidth]{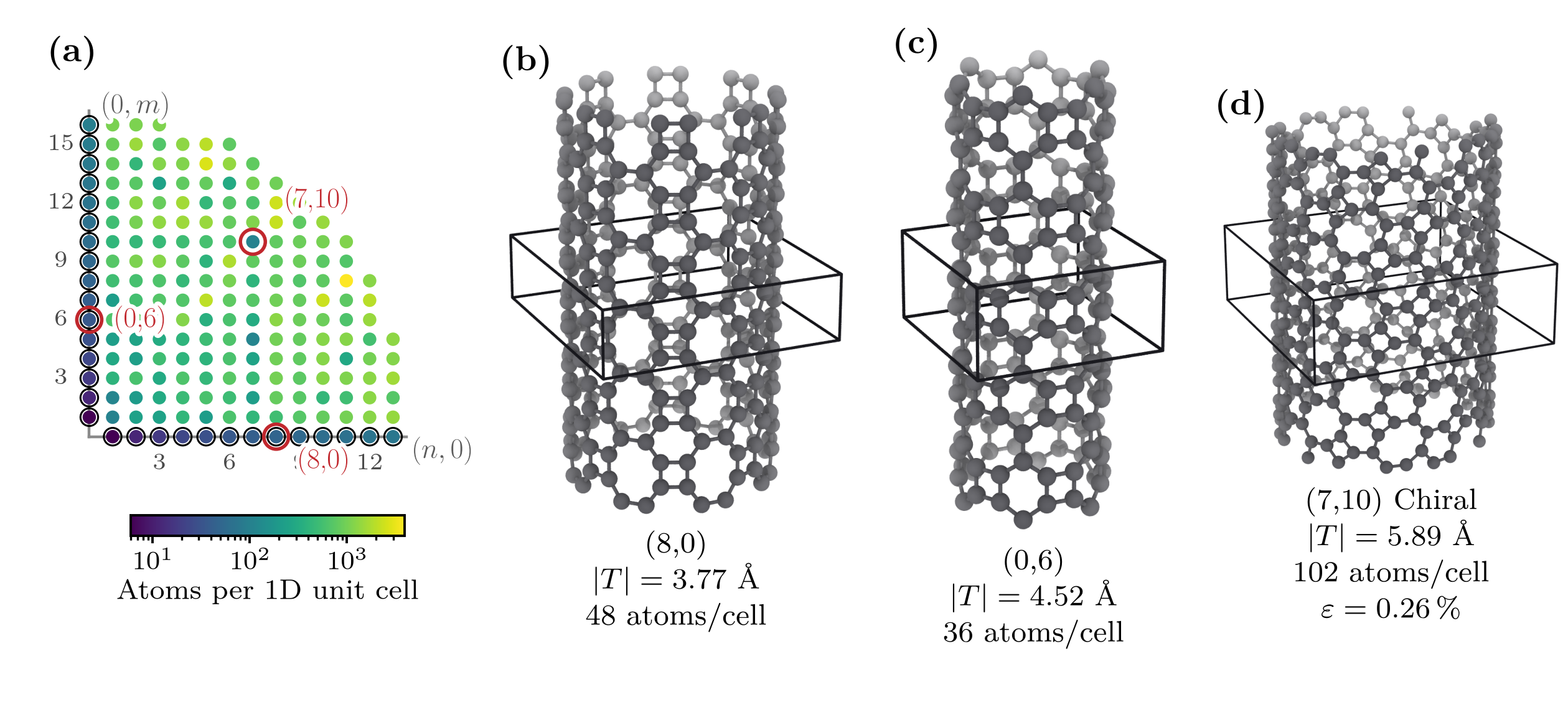}
  \caption{Chirality map of the biphenylene network up to $D = 20$~\AA, colored by the number of atoms in the smallest cell with $\varepsilon \le 0.5\%$, with exact cells ringed (a), and the $(8,0)$, $(0,6)$, and $(7,10)$ tubes with one unit cell highlighted (b--d).}
  \label{fig:biphenylene}
\end{figure*}

The biphenylene network\cite{Fan2021} displays the opposite regime, since for its rectangular cell, taken from the database of Shi \textit{et al.},\cite{Shi2021} Eq.~\ref{eq:rectangular} predicts that only the two axial families $(n,0)$ and $(0,m)$, for which $\T$ is a lattice vector, close with small cells. Figure~\ref{fig:biphenylene} shows its $90^\circ$ chirality map up to $D = 20$~\AA, colored by the number of atoms in the smallest cell with a residual of at most 0.5\%.

\begin{figure}[!t]
  \centering
  \includegraphics[width=\linewidth]{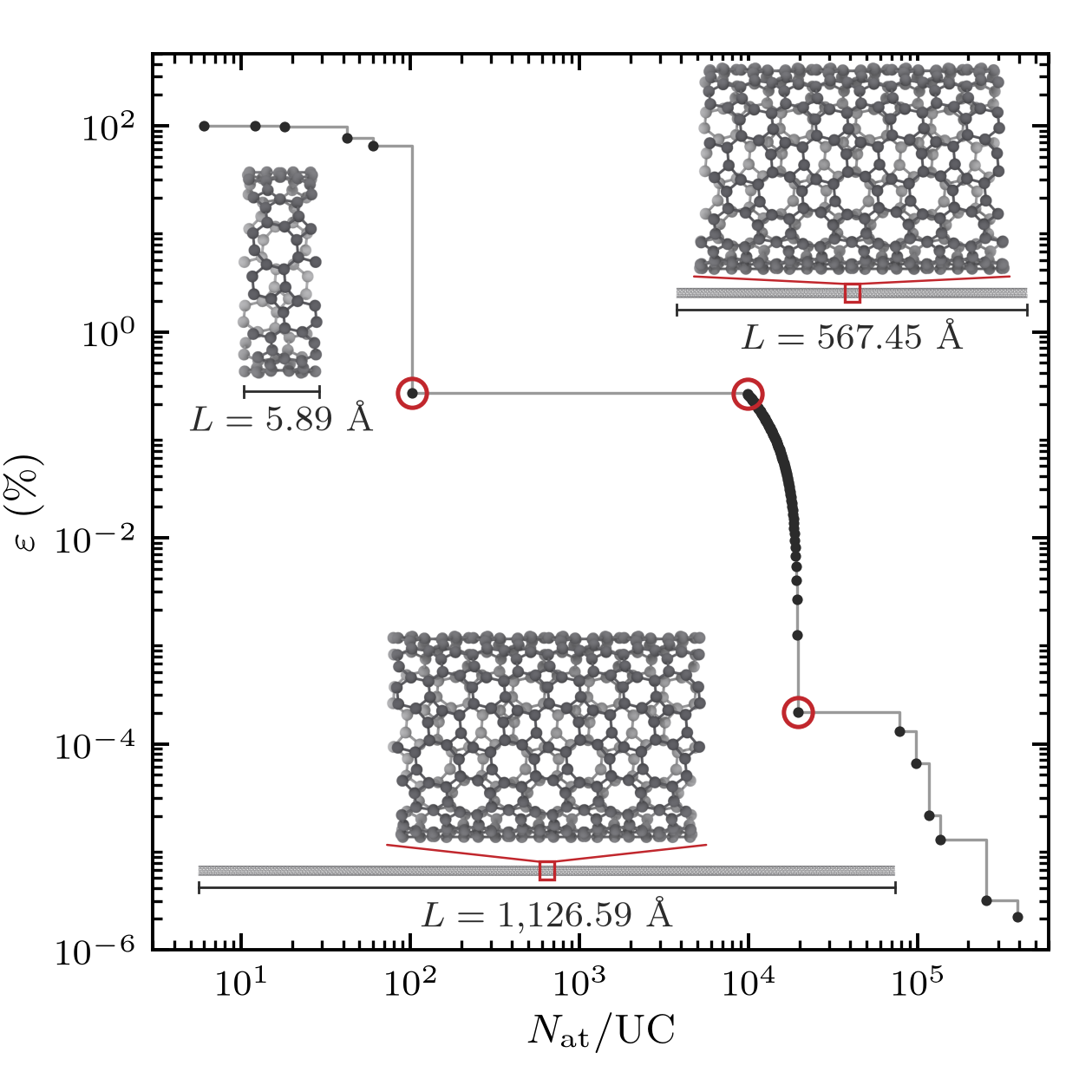}
  \caption{Periodicity residual $\varepsilon$ as a function of the number of atoms per cell along the trade-off front of the $(7,10)$ tube of the biphenylene network, with the unit cells of the sixth, seventh, and 102nd cells of the front.}
  \label{fig:front}
\end{figure}

Of the 195 index pairs of the map, 29 close exactly, and all of them belong to the axial families. The $(8,0)$ and $(0,6)$ tubes, with diameters of 11.51 and 7.20~\AA, have periods equal to the lattice constants of 3.77 and 4.52~\AA\ and cells of 48 and 36 atoms. The chiral $(7,10)$ tube, with $D = 15.67$~\AA, is built with $\T = -\av + \bv$ in a cell of 102 atoms and 5.89~\AA\ that leaves a residual of 0.26\%, whereas the $(4,12)$ tube, at almost the same diameter of 15.51~\AA, needs 1,584 atoms and 92.33~\AA\ to reach a residual below 0.5\%. Two tubes of nearly equal size can thus differ by more than an order of magnitude in cost, because the rational approximations of their ratios $\alpha$ converge at different rates. Figure~\ref{fig:front} follows the $(7,10)$ tube along its front.

Within a search bound of 5,000, the front of the $(7,10)$ tube contains 108 cells. The sixth, with 102 atoms and a residual of 0.26\%, is the first cell of practical use. The seventh opens a dense sequence of approximations with $\T = -j\,\av + (j+1)\,\bv$, which grow from 9,834 atoms and 567.45~\AA\ at $j = 96$ to 19,524 atoms and 1,126.59~\AA\ at $j = 191$, while the residual decreases from 0.25\% to $3.9\times10^{-3}$\% at $j = 188$ and reaches $2.0\times10^{-4}$\% only at the last cell of the sequence. These cells illustrate Eq.~\ref{eq:bound}, since the sixth cell, a convergent of $\alpha$, lies far below its bound of $S/|\T|^2 = 49\%$, whereas the cells of the dense sequence are intermediate fractions, which the bound does not cover, and exceed it up to $j = 189$, while the cells with $j = 190$ and $j = 191$ are convergents and satisfy it, the last with $S/|\T|^2 = 1.3\times10^{-3}$\%. The exact cell of the stored lattice constants lies far beyond the front, with 2,422,690,789 primitive cells, $1.5\times10^{10}$ atoms, and a length of 8.4~cm. A cell suitable for first-principles calculations and a cell periodic to machine precision can therefore be separated by eight orders of magnitude in size, and Eq.~\ref{eq:front} turns this choice into an explicit selection along the front.

\subsection{Symmetry and Rolling Sense}
\label{sec:symmetry}

The sectors of $30^\circ$ and $90^\circ$ used above are those of the lattices, and they are correct only when the atoms preserve the symmetry of their lattice. In general, an operation $R$ of the point group of the decorated layer maps lattice vectors onto lattice vectors through an integer matrix $U_R$, and the tube built from $(n,m)$ is congruent to the tube built from $(n,m)\,U_R$. A proper operation, with $\det U_R = 1$, gives the same tube, and an improper one, with $\det U_R = -1$, gives its mirror image, which has the same diameter, period, number of atoms, and energy and is counted once. Operations that keep the two faces of the layer in place preserve the rolling sense $\sigma$ of Eq.~\ref{eq:rolling}, while operations that exchange the faces, such as a horizontal mirror, a rotoreflection, or a twofold axis lying in the plane, relate the tube of $(n,m)$ rolled with $\sigma$ to the tube of $(n,m)\,U_R$ rolled with $-\sigma$. The chirality map of a crystal is therefore the quotient of the index lattice by a group $G$ of integer matrices, and its irreducible sector spans $360^\circ/|G|$.

The 17 plane groups\cite{Aroyo2016} collapse onto far fewer maps for two reasons. A glide reflection acts on the rolled tube as its mirror component followed by a translation of the sheet, and a translation of the sheet only shifts the origin of the tube, so that $G$ depends only on the point group and on its orientation with respect to the lattice, which is the arithmetic crystal class, and the 17 groups reduce to 13 classes. In addition, $(n,m)$ and $(-n,-m)$ always give the same tube, since reversing the chiral vector rolls the same strip around the same axis. The matrix $-I$ is the twofold rotation about the normal of the layer, which keeps both faces in place, so rolling adds this rotation to every point group. Each of the 13 classes then merges with its centrosymmetric partner among $p2$, $p2mm$, $c2mm$, $p4$, $p4mm$, $p6$, and $p6mm$, and the 17 plane groups induce exactly seven distinct chirality maps. Table~\ref{tab:classes} lists them, with the counts of distinct tubes obtained from a representative structure built for each plane group.

\begin{table}[!t]
  \centering
  \caption{Classes of plane groups that induce the same chirality map, with the order $|G|$ of the group acting on the chiral indices, the width of the irreducible sector, and the number of distinct tubes with $|n|, |m| \le 6$.}
    \label{tab:classes}
  \small
  \setlength{\tabcolsep}{3pt}
  \begin{tabular}{lrrr}
    \toprule
    Plane groups & $|G|$ & Sector ($^\circ$) & Tubes \\
    \midrule
    $p1$, $p2$                          &  2 & 180 & 84 \\
    $pm$, $pg$, $pmm$, $pmg$, $pgg$     &  4 &  90 & 48 \\
    $cm$, $cmm$                         &  4 &  90 & 48 \\
    $p4$                                &  4 &  90 & 42 \\
    $p4m$, $p4g$                        &  8 &  45 & 27 \\
    $p3$, $p6$                          &  6 &  60 & 42 \\
    $p3m1$, $p31m$, $p6m$               & 12 &  30 & 27 \\
    \bottomrule
  \end{tabular}
\end{table}

\begin{figure*}[!t]
  \centering
  \includegraphics[width=\linewidth]{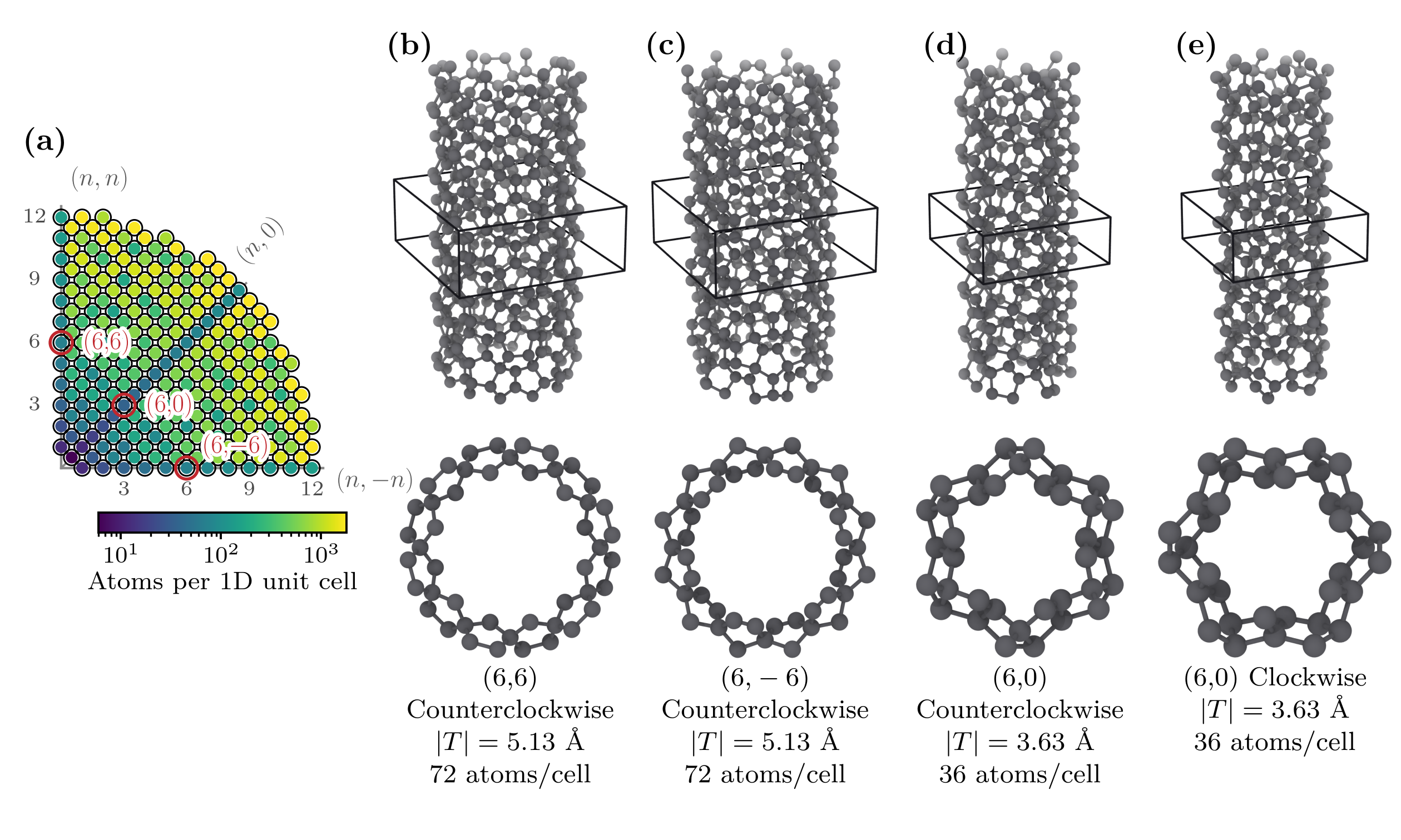}
  \caption{Chirality map of penta-graphene between $(n,-n)$ and $(n,n)$, with the $(n,0)$ diagonal dashed (a), the $(6,6)$ (b) and $(6,-6)$ (c) tubes rolled counterclockwise, and the $(6,0)$ tube rolled counterclockwise (d) and clockwise (e), each with its cross section.}
  \label{fig:penta}
\end{figure*}

The rectangular and centered rectangular classes of Table~\ref{tab:classes} share the same order and the same tube count but not the same set of tubes, because the mirror lies along the lattice vectors in the first case and along their bisector in the second. The classification assumes that the operations of the layer keep its faces in place, and penta-graphene, whose nanotubes have been studied by first-principles and molecular dynamics simulations,\cite{Chen2017, QuijanoBriones2017, Wang2017} provides an example in which they do not. Its space group $P\bar{4}2_1m$\cite{Zhang2015} projects onto the plane as the square group $p4gm$, which would suggest a $45^\circ$ sector, but its fourfold operation is a rotoreflection that exchanges the two faces of the buckled layer. Figure~\ref{fig:penta} shows the resulting chirality map together with tubes rolled in both senses.

For a fixed rolling sense, the group acting on the indices of penta-graphene contains only four operations, namely the identity, the rotation $-I$ supplied by rolling, the exchange $(n,m) \to (m,n)$ of its diagonal glide, which yields mirror-image tubes, and their product. The irreducible map is therefore the $90^\circ$ wedge between $(n,-n)$ and $(n,n)$ shown in Figure~\ref{fig:penta}(a). The $(6,6)$ and $(6,-6)$ tubes of panels (b) and (c) have the same diameter of 9.80~\AA, the same period of 5.13~\AA, and the same 72 atoms per cell, but they are different tubes in the same sense and become identical only when the sense of one of them is inverted, because the rotoreflection maps $(n,m)$ onto $(-m,n)$ while exchanging the faces. On the $(n,0)$ line, the twofold screw axes lying in the plane of the layer relate the two senses by a reflection, so that the $(6,0)$ tubes of panels (d) and (e), with $D = 6.93$~\AA\ and 36 atoms, are mirror images of each other. The $45^\circ$ half-wedge between $(n,0)$ and $(n,n)$ thus contains every tube only if both senses are used, and for penta-graphene the choice of chirality and the choice of rolling sense are not independent.

\subsection{Curvature and Bond Changes}
\label{sec:curvature}

\begin{figure*}[!t]
  \centering
  \includegraphics[width=0.8\linewidth]{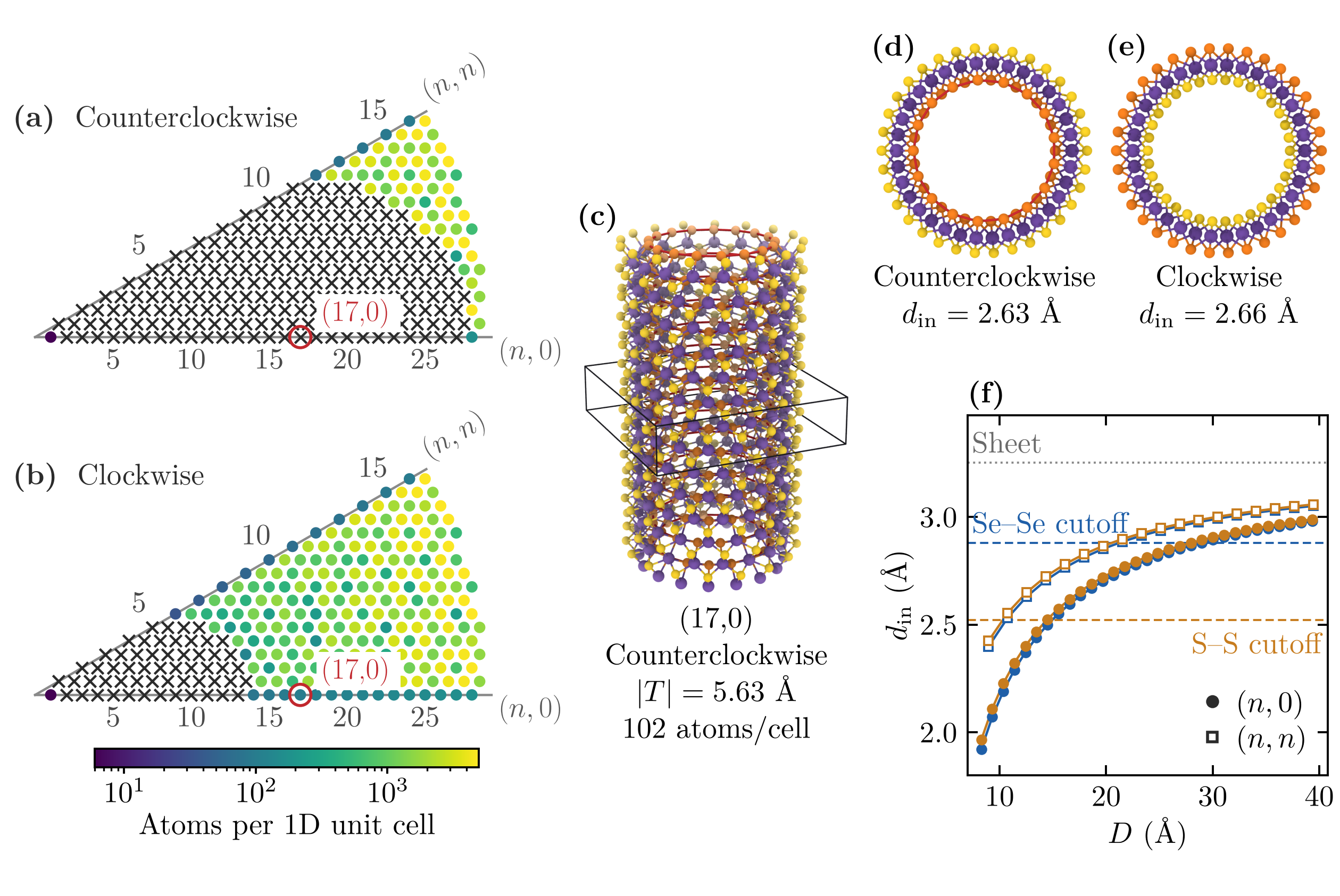}
  \caption{Chirality maps of \ce{MoSSe} rolled counterclockwise with Se inside (a) and clockwise with S inside (b), with crosses on tubes with bonds formed or broken by rolling, the $(17,0)$ tube rolled counterclockwise with the formed bonds in red (c), cross sections of the $(17,0)$ tube in both senses (d, e), and the shortest distance in the inner chalcogen ring as a function of the diameter (f).}
  \label{fig:mosse}
\end{figure*}

Janus \ce{MoSSe} removes the last relation between the two senses, since its layer has neither a horizontal mirror nor any operation that exchanges its S and Se faces.\cite{Lu2017, Zhang2017} Its two rolling senses therefore define two complete sets of tubes, one with Se on the inner wall and one with S,\cite{Luo2019, Evarestov2020} whose relative stability depends on which chalcogen lies inside,\cite{Bolle2021} and they differ both in which tubes exist and in which tubes preserve the bonding of the layer. Eq.~\ref{eq:rolling} scales the arc length between two neighbors at height $\zeta$ by $(R + \sigma\zeta)/R$, so a thick layer compresses its inner face and stretches its outer face, and the resulting distances can cross the thresholds that define a bond. We detect these changes pair by pair, using the fact that every atom of the tube keeps its identity in the flat layer, so that each pair of atoms $i$ and $j$ has a distance $d^{\,0}_{ij}$ in the sheet and a distance $d_{ij}$ in the tube, and both are compared with the cutoff
\begin{equation}
  d^{\,c}_{ij} = 1.2\,(r_i + r_j),
  \label{eq:cutoff}
\end{equation}
\noindent where $r_i$ and $r_j$ are the covalent radii of Cordero \textit{et al.}\cite{Cordero2008} A bond is formed by rolling when $d_{ij} < d^{\,c}_{ij} \le d^{\,0}_{ij}$ and $d_{ij} \le 0.9\,d^{\,0}_{ij}$, and it is broken when $d^{\,0}_{ij} < d^{\,c}_{ij} \le d_{ij}$ and $d_{ij} \ge 1.1\,d^{\,0}_{ij}$. The 10\% margin separates a change of bonding from the slight compression or extension that rolling imposes even on wide tubes, which would otherwise flag tubes of any diameter whenever a pair of the sheet lies close to its cutoff. Figure~\ref{fig:mosse} applies the test to the chirality map of \ce{MoSSe} in both senses.

Of the 276 tubes in each map of Figure~\ref{fig:mosse}(a) and (b), 215 form new bonds with Se inside, with affected tubes up to $D = 28.6$~\AA, whereas only 56 are affected with S inside, none beyond 14.01~\AA, and no tube of either map breaks a bond of the layer. The $(17,0)$ tube of panel (c), with $D = 17.6$~\AA\ and 102 atoms, illustrates the Se-inside case, in which the new bonds join neighboring Se atoms around the inner ring. Panels (d) to (f) show that the asymmetry does not arise from a different compression. The shortest distance in the inner ring of the $(17,0)$ tube is 2.63~\AA\ for Se inside and 2.66~\AA\ for S inside, starting from the same in-plane distance of 3.25~\AA. What differs is the cutoff of Eq.~\ref{eq:cutoff}, 2.88~\AA\ for Se-Se and 2.52~\AA\ for S-S, since the covalent radius of Se exceeds that of S. For two inner atoms aligned with the circumference, as in the zigzag tubes, the inner face lies at $h/2$ below the mid-plane, with $h$ the distance between the two chalcogen planes, and Eq.~\ref{eq:rolling} compresses their separation $a$ to $a\,(1 - h/D)$. Setting this distance equal to the cutoff $d^{\,c}_{\mathrm{in}}$ of Eq.~\ref{eq:cutoff} for the pair of inner atoms gives the critical diameter
\begin{equation}
  D_c = \frac{h\,a}{a - d^{\,c}_{\mathrm{in}}},
  \label{eq:dcrit}
\end{equation}
\noindent which, with $a = 3.253$~\AA\ and $h = 3.206$~\AA, predicts 28.0 and 14.2~\AA\ for Se and S inside, in close agreement with the largest affected diameters of 28.60 and 14.01~\AA\ in the two maps. Eq.~\ref{eq:dcrit} makes explicit that bond changes induced by curvature depend on the thickness of the layer and on the chemistry of its inner face. Penta-graphene, with a buckling of 1.37~\AA, lies at the opposite limit, since only 11 of the 249 index pairs of its map show bond changes in either sense, all of them with $D < 3.7$~\AA. The tubes that pass the test keep the bonding pattern of the flat layer and are the natural starting geometries for relaxation and for mechanical loading, under which the inner shell of penta-graphene tubes forms bonds between neighboring carbon dimers at moderate axial compression.\cite{Wang2017}

\section{Structures Built from the Tube}
\label{sec:beyond}

The tubes of Section~\ref{sec:rolling} are the building blocks of more complex structures, and each of these structures poses its own condition of periodicity. The following subsections treat the deformation of a single tube, multiwalled tubes, and bundles, and show that these operations can be combined on the same structure.

\subsection{Strain and Torsion}
\label{sec:deformations}

A rolled tube is the starting point for structures that probe its mechanical response, and the simplest of them deform a single cell. Axial strain maps every coordinate along the axis as $z \to z\,(1 + \epsilon_z)$ together with the length of the cell, which preserves periodicity for any $\epsilon_z$. Torsion rotates each atom about the axis by
\begin{equation}
  \phi(z) = \frac{\Phi\, z}{L},
  \label{eq:torsion}
\end{equation}
\noindent where $\Phi$ is the total twist accumulated over the cell length $L$, and the lower band of Figure~\ref{fig:roll} shows both operations on the $(8,0)$ tube of graphene. Unlike strain, torsion keeps the cell periodic only if the rotation accumulated over one period maps the tube onto itself. A tube rolled from a primitive layer with $\Ch = \nu\,\Ch'$ wraps $\nu$ copies of the primitive strip around its circumference, so it is invariant under rotations by $360^\circ/\nu$ about its axis, and a periodic twisted cell requires $\Phi$ to be a multiple of that angle. \ntbuilder{} tests this condition directly, by checking whether the rotation by $\Phi$ maps the untwisted structure onto itself, keeps the axial period of a closing twist, and returns any other twist as a finite segment with vacuum along the axis. Helical structures can also be generated without a translational cell, as objective structures built from a screw operation,\cite{James2006, Dumitrica2007} and the closing condition is what allows twisted tubes and bundles to enter codes with periodic boundary conditions. The deformation imposed by a rigid twist follows from its geometry, since at a distance $r$ from the axis Eq.~\ref{eq:torsion} is a simple shear of magnitude $\gamma_s = \Phi r/L$ along the circumference, whose largest principal stretch is
\begin{equation}
  \lambda_{\max} = \frac{\gamma_s}{2} + \sqrt{1 + \frac{\gamma_s^{2}}{4}},
  \label{eq:stretch}
\end{equation}
\noindent attained along the direction at an angle $\arctan(1/\lambda_{\max})$ from the axis, which tends to $45^\circ$ for small twists. Bonds along the circumference keep their length, whereas bonds close to that direction elongate by up to $\lambda_{\max} - 1$, a value that grows linearly with $r$ for small twists and sets how far a twisted cell is from a relaxed structure.

\subsection{Multiwalled Tubes}
\label{sec:mwnt}

\begin{figure}[!b]
  \centering
  \includegraphics[width=\linewidth]{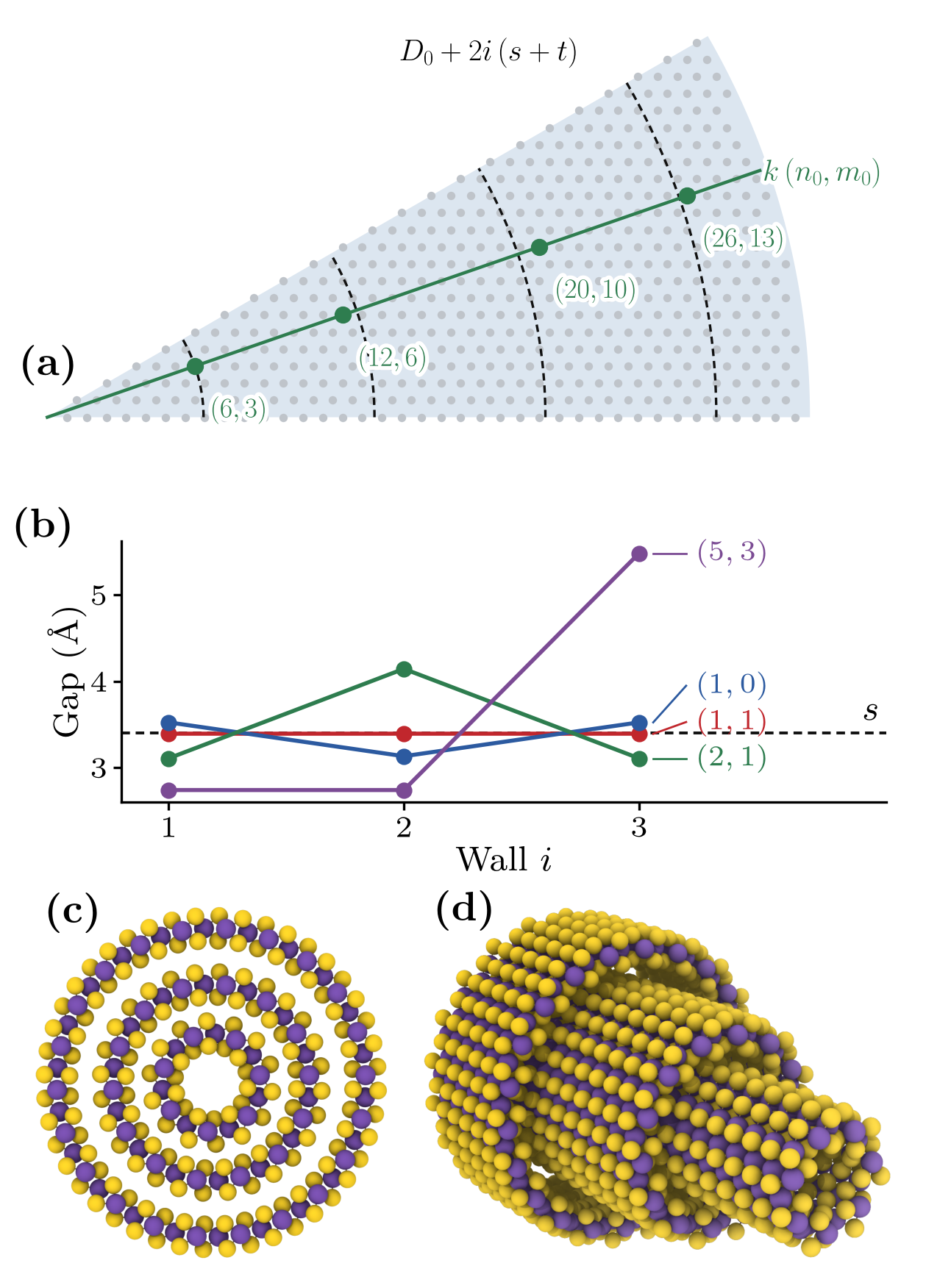}
  \caption{Chiral vectors of graphene with the target diameters (dashed arcs) and the walls chosen along the ray of the $(6,3)$ family (a), realized gap as a function of the wall index for four families with the requested gap $s$ (b), and the \ce{MoS2} $(7,7)@(14,14)@(22,22)$ tube viewed along its axis (c) and with walls of different lengths (d).}
  \label{fig:mwnt}
\end{figure}

Adding walls around a tube raises the same question of periodicity as torsion, since a multiwalled tube is periodic only if all its walls share an axial period, and Eq.~\ref{eq:costlaw} shows how to guarantee it. The walls $k\,(n_0,m_0)$, with $(n_0,m_0) = (n,m)/\gcd(n,m)$, have the same ratio $r$ of Eq.~\ref{eq:ratio} as the inner tube $(n,m)$ and therefore the same translation vector, the same chiral angle, and the same period, whereas walls taken from other families share a period only when the ratio of their periods is rational, and then only in a common multiple of both. Commensurate walls are a modeling choice, since double-walled carbon tubes observed in experiment combine walls whose chiralities are not correlated,\cite{Hirahara2006} while commensurate walls provide the common periodic cell used in first-principles studies of multiwalled carbon, BN, and \ce{TiO2} tubes.\cite{Charlier1993, Evarestov2011} \ntbuilder{} places the walls on the ray of the inner tube, with the target diameter of wall $i$ given by
\begin{equation}
  D_i^{\ast} = D_0 + 2i\,(s + t),
  \label{eq:mwnt_target}
\end{equation}
\noindent where $D_0$ is the diameter of the inner tube, $s$ the requested gap between walls, and $t$ the thickness of the layer, and the multiple $k_i$ is the integer closest to $D_i^{\ast}/d_0$, with $d_0 = D_0/\gcd(n,m)$ the diameter of the primitive tube of the family, constrained to increase with $i$. The realized gap
\begin{equation}
  s_i = \frac{D_i - D_{i-1}}{2} - t = \frac{(k_i - k_{i-1})\,d_0}{2} - t
  \label{eq:mwnt_gap}
\end{equation}
\noindent is quantized in steps of $d_0/2$, so exact periodicity comes at the expense of a coarser control of the spacing. Figure~\ref{fig:mwnt} illustrates the choice of walls and its consequences.

For graphene, $t = 0$ and a requested gap of 3.4~\AA, the inner $(6,3)$ tube receives the walls $(12,6)$, $(20,10)$, and $(26,13)$ of Figure~\ref{fig:mwnt}(a). Figure~\ref{fig:mwnt}(b) shows how the quantization of Eq.~\ref{eq:mwnt_gap} depends on the family. The armchair family $(1,1)$, with a step of 0.68~\AA, realizes gaps of 3.39~\AA\ for every wall, the zigzag family $(1,0)$ alternates between 3.52 and 3.13~\AA, the $(2,1)$ family between 3.11 and 4.14~\AA, and the $(5,3)$ family, with a step of 2.74~\AA, allows only gaps of 2.74 or 5.48~\AA. The larger the primitive vector of the family, the coarser the available spacings.

The thickness term of Eq.~\ref{eq:mwnt_target} becomes essential for non-planar layers. For \ce{MoS2}, with $t = 3.13$~\AA\ and a primitive diameter of 1.76~\AA\ in the armchair family, a three-walled tube built on $(7,7)$ requires targets of 25.38 and 38.43~\AA, which select $k = 14$ and $k = 22$ and give the $(7,7)@(14,14)@(22,22)$ tube of Figure~\ref{fig:mwnt}(c) and (d), with 258 atoms per period of 3.19~\AA\ and gaps of 3.03 and 3.91~\AA\ between the sulfur surfaces of consecutive walls. All three walls are free of curvature-induced bond changes, whereas the smaller inner tube $(6,6)$ forms S-S bonds on its inner face, so that the criterion of Section~\ref{sec:curvature} also guides the choice of the inner wall.

\subsection{Bundles and Combined Operations}
\label{sec:bundles}

\begin{figure}[!t]
  \centering
  \includegraphics[width=\linewidth]{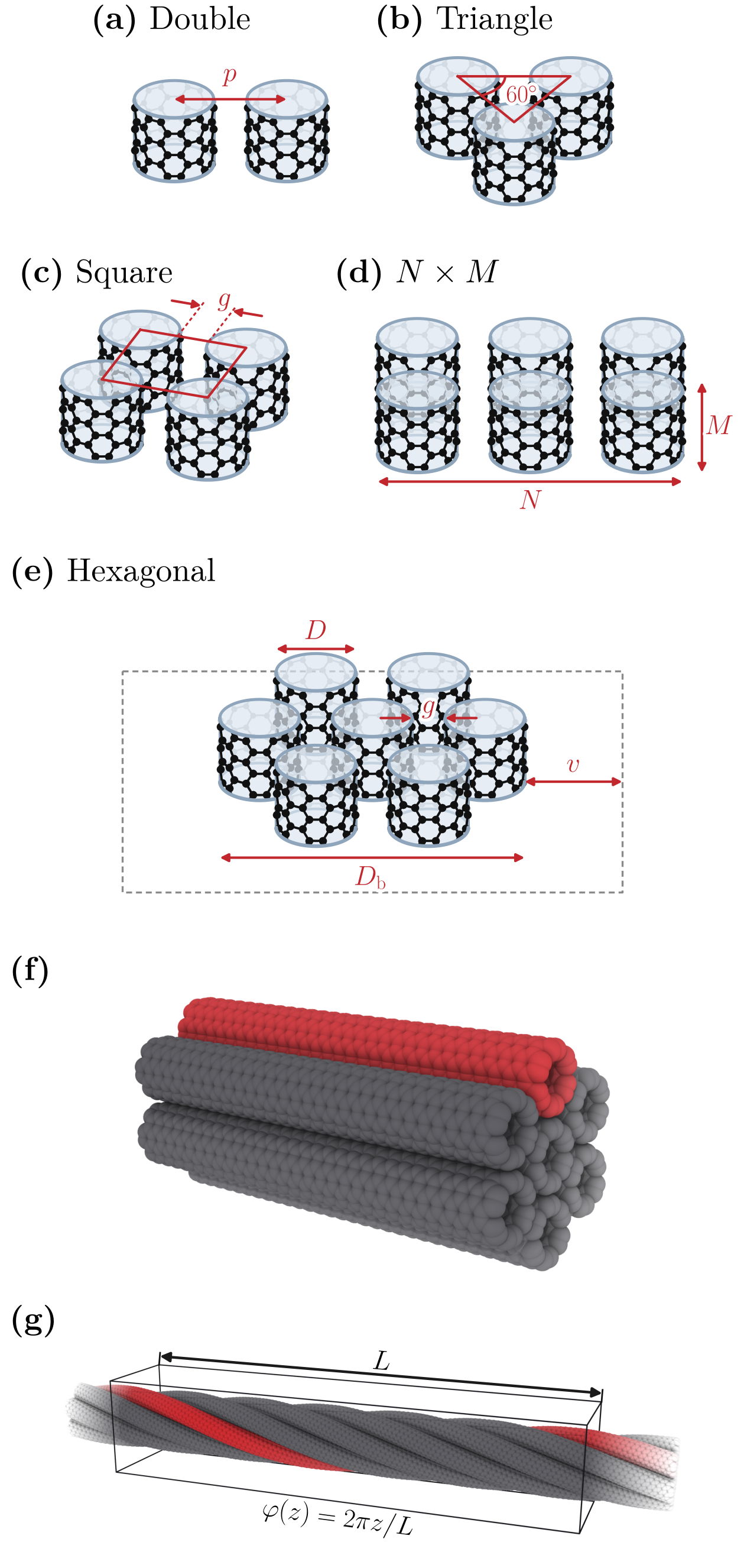}
  \caption{Bundle lattices with their parameters, Double (a), Triangle (b), Square (c), $N \times M$ (d), and Hexagonal (e), a straight hexagonal bundle of seven $(6,6)$ tubes (f), and the same bundle replicated along the axis and twisted by $360^\circ$ over its length (g).}
  \label{fig:bundles}
\end{figure}

Experimentally grown nanotubes often assemble into bundles,\cite{Thess1996} held together by van der Waals interactions,\cite{Tersoff1994} which adds a second lattice, transverse to the axis, to the construction. \ntbuilder{} replicates a tube on a 2D lattice of pitch $p = D_{\mathrm{out}} + g$, where $D_{\mathrm{out}}$ is the outer diameter measured from the atoms and $g$ the gap between tube surfaces, and all the operations of this section can be chained on the same structure. Figure~\ref{fig:bundles} shows the available bundle lattices and one such combination.

The Triangle and Hexagonal packings of Figure~\ref{fig:bundles}(b) and (e) build finite clusters of three and seven tubes inside a supercell with a vacuum layer, which suits the study of small bundles and of the interaction between neighboring tubes. The Double, Square, and $N \times M$ packings of panels (a), (c), and (d) place the tubes on a rectangular grid, which becomes a periodic array of tubes when the vacuum equals half the gap, since the periodic images then lie at the same pitch as the tubes inside the cell. For a twisted bundle, the condition of Section~\ref{sec:deformations} involves both lattices. A rotation of the hexagonal bundle by $60^\circ$ about its central axis carries each outer tube onto the position of its neighbor while rotating it by $60^\circ$ about its own axis, so the smallest closing twist must be a multiple of both $60^\circ$ and $360^\circ/\gcd(n,m)$. For $(6,6)$ tubes this angle is $60^\circ$, whereas a $(5,5)$ bundle twisted by $60^\circ$ does not close, because its fivefold axis is incompatible with the sixfold lattice and the smallest closing twist is $360^\circ$.

Figure~\ref{fig:bundles}(f) and (g) chain these operations, starting from seven $(6,6)$ tubes with $g = 3.4$~\AA\ arranged in a hexagonal bundle, which is replicated along the axis to 102 cells and a length of 251~\AA\ and twisted by $360^\circ$ over that length, a multiple of the $60^\circ$ closing angle, producing a helical fiber that is exactly periodic along its axis. For the outermost atoms of the outer tubes, at about 15.6~\AA\ from the axis of the bundle, the shear is $\gamma_s \approx 0.39$ and Eq.~\ref{eq:stretch} gives a largest stretch of about 21\%, consistent with the bond elongations of up to 20\% measured in the twisted cell. Such structures are starting geometries for relaxation, and the exact periodicity of the cell makes them directly usable in periodic simulations.

\section{Software and Catalog}
\label{sec:software}

The constructions of Sections~\ref{sec:rolling} and \ref{sec:beyond} are implemented in \ntbuilder{}, whose organization and interfaces are described first, followed by their application to a catalog of nanotubes built from public databases of 2D materials.

\subsection{Implementation and Interfaces}
\label{sec:implementation}

\begin{figure*}[!t]
  \centering
  \includegraphics[width=0.7\linewidth]{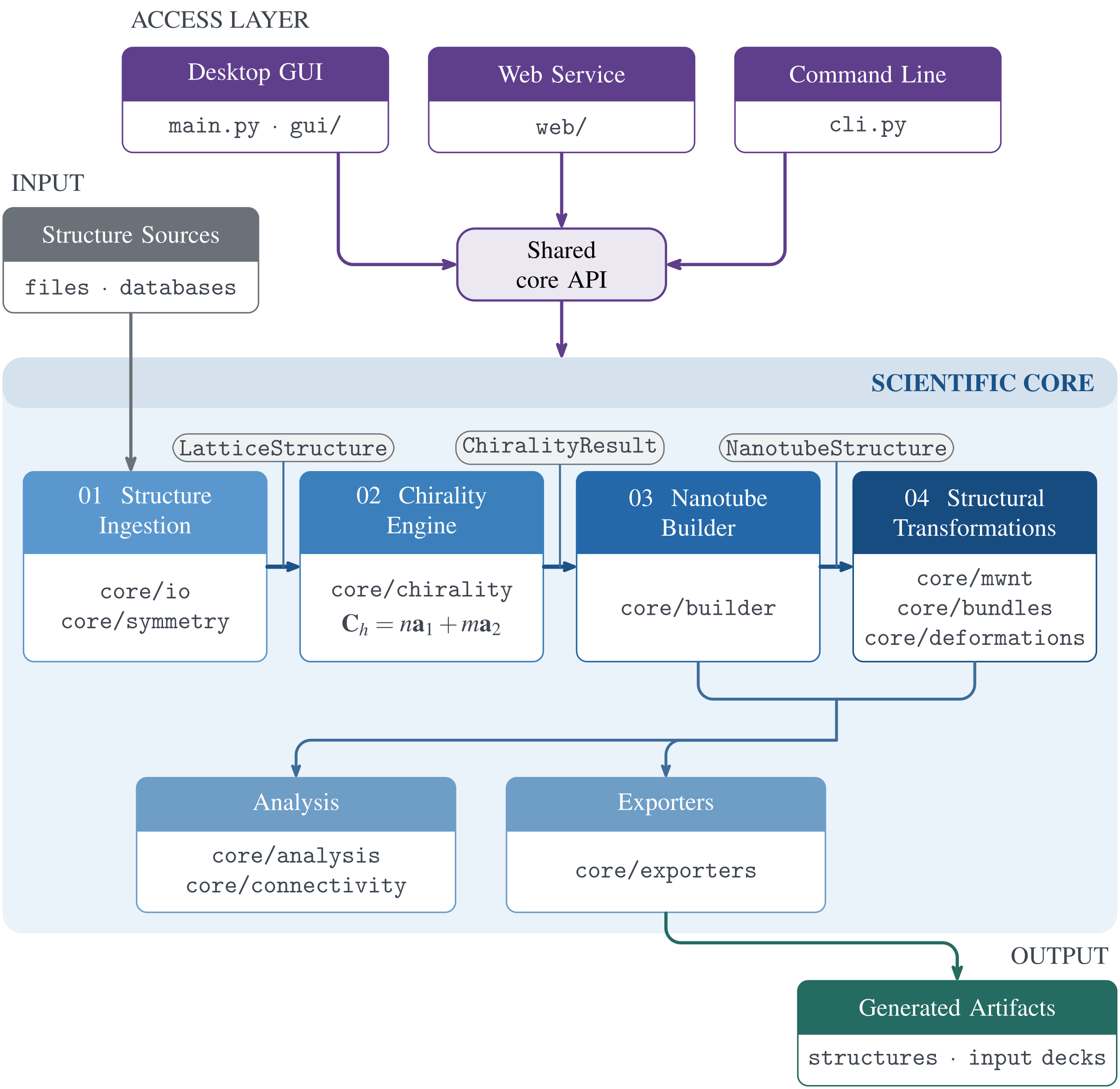}
  \caption{Architecture of \ntbuilder{}, with the interfaces (top), the modules of the core library, and the data passed between them from the input structure to the exported files.}
  \label{fig:architecture}
\end{figure*}

The core of \ntbuilder{} is a Python library that does not depend on any graphical component, with two interfaces and a command-line entry point built on top of it. Figure~\ref{fig:architecture} shows the organization of the code and the flow of data from the input structure to the exported files.

\begin{figure*}[!t]
  \centering
  \includegraphics[width=0.7\linewidth]{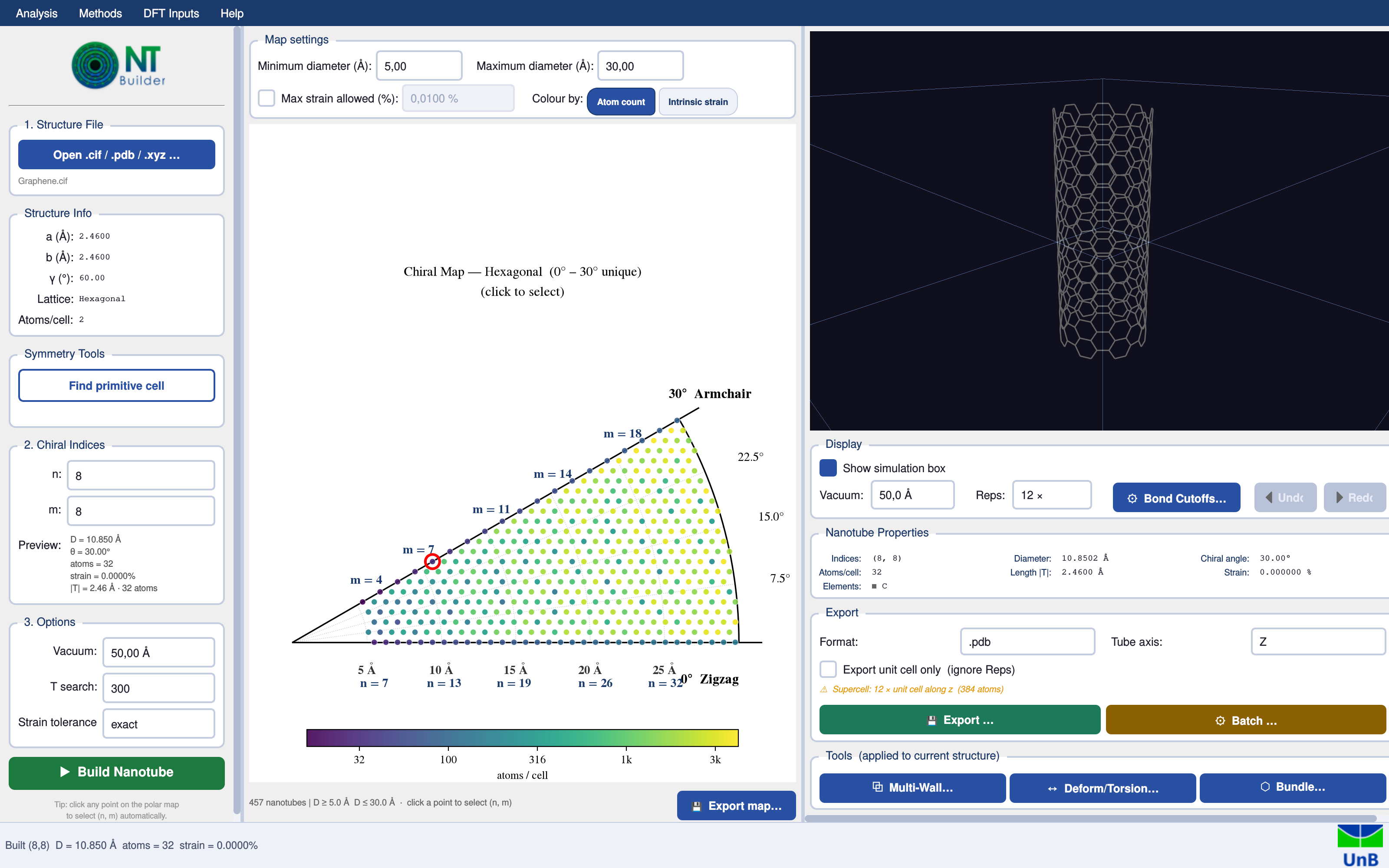}
  \caption{Desktop interface of \ntbuilder{} with the graphene layer loaded, its chirality map, and the $(8,8)$ tube in the 3D viewer.}
  \label{fig:desktop}
\end{figure*}

\begin{figure*}[!t]
  \centering
  \includegraphics[width=0.7\linewidth]{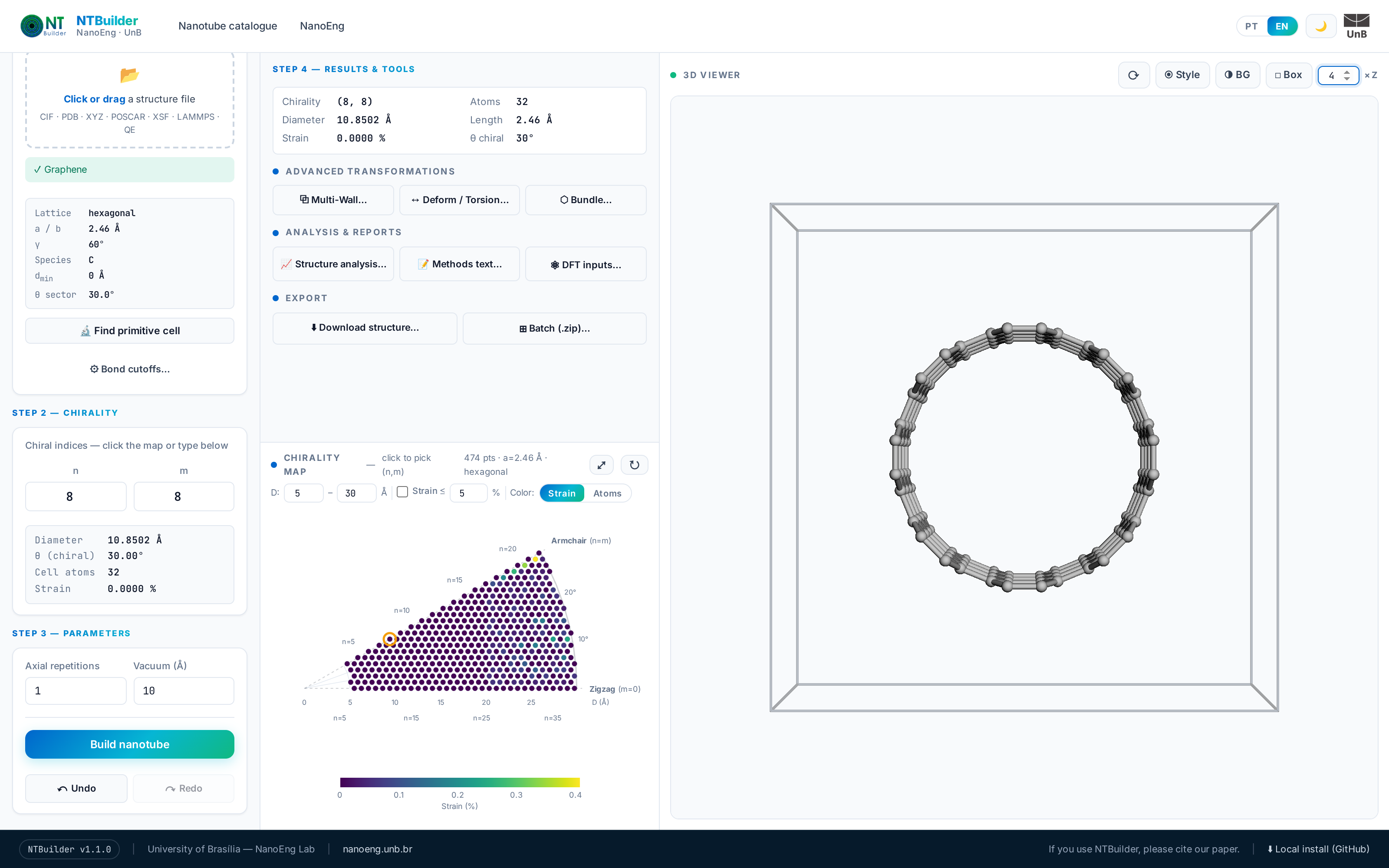}
  \caption{Web interface of \ntbuilder{} with the graphene layer loaded, its chirality map, and the $(8,8)$ tube in the 3D viewer.}
  \label{fig:web}
\end{figure*}

Structures are read from CIF, PDB, XYZ, POSCAR, XSF, LAMMPS, and Quantum ESPRESSO files, or retrieved from the Crystallography Open Database, the Materials Project, and C2DB.\cite{Grazulis2012, Jain2013, Haastrup2018} CIF files are parsed with gemmi, with a built-in parser as fallback,\cite{Wojdyr2022} after which a single layer is isolated and oriented in the $xy$ plane. The lattice is then snapped to its ideal Bravais class with tolerances of at most 2\% on the lengths and $1.2^\circ$ on the angle, and the point group of the decorated layer is detected with an absolute tolerance on atomic positions chosen from the sequence 0.001, 0.01, 0.1, and 0.5~\AA\ as the tightest value that yields the largest group, following the tolerance ladders used by crystallographic databases. The group acting on the indices, the irreducible sector, and the list of distinct tubes follow from the rules of Section~\ref{sec:symmetry}.

\begin{figure*}[!t]
  \centering
  \includegraphics[width=\linewidth]{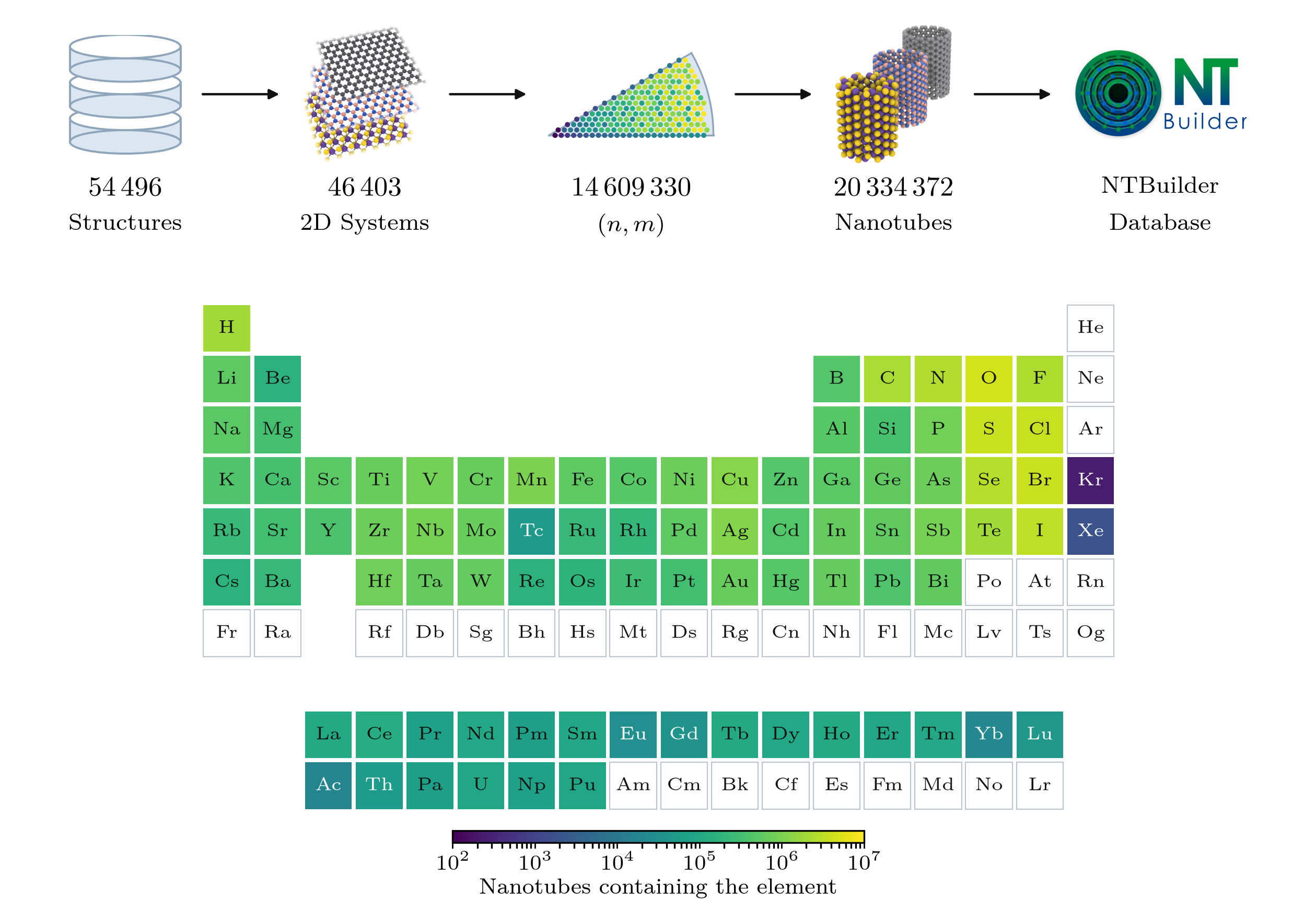}
  \caption{Workflow of the nanotube catalog from the source databases to the 2D systems, chiral indices, and nanotubes (top), and number of nanotubes containing each element (bottom).}
  \label{fig:catalog}
\end{figure*}

The default search for the translation vector visits the convergents of $\alpha$ in Eq.~\ref{eq:alpha} with one intermediate fraction per level, so that its cost grows logarithmically with the bound, and the complete front is enumerated lazily when a residual tolerance or a maximum length is requested. The builder places the atoms of the strip spanned by $\Ch$ and $\T$ with an iterator over analytic bounds that never materializes the supercell of the sheet, which keeps the memory proportional to the number of output atoms even for cells of millions of atoms, and it applies Eq.~\ref{eq:rolling} in either sense together with the bond test of Eq.~\ref{eq:cutoff}, with covalent radii for the elements from H to Cm.\cite{Cordero2008} The resulting structures can be converted into multiwalled tubes, bundles, and strained or twisted cells, analyzed in terms of bond statistics and line-group metadata, accompanied by a Methods paragraph and input decks for VASP, Quantum ESPRESSO, and CP2K, and exported as POSCAR, Quantum ESPRESSO, LAMMPS, XYZ, PDB, XSF, CP2K, SIESTA, and CIF files, either for one tube or for a whole family of indices.\cite{Kresse1996, Giannozzi2009, Kuhne2020, Plimpton1995, Soler2002} The library is covered by 126 automated tests.

The same library drives a desktop application, built with PyQt6 and an OpenGL viewer based on pyqtgraph,\cite{RiverbankComputing2021, Moore2023} and a web service, served by FastAPI with an interactive chirality map and a 3Dmol viewer,\cite{Ramirez2018, Rego2015} so that the full workflow runs in a browser without installation. Figures~\ref{fig:desktop} and \ref{fig:web} show both interfaces in the same state, with the graphene layer loaded, its chirality map computed, and the $(8,8)$ tube built.

In both interfaces the workflow follows the order of this work, in which the user loads a layer, selects a tube on the chirality map or by its indices, chooses the rolling sense and the cell, and applies the operations of Section~\ref{sec:beyond} before exporting. In the web interface, the curvature test runs in the background after the map is displayed and marks the affected tubes as it progresses, so that even layers with large cells remain interactive.

\subsection{A Catalog of Nanotubes}
\label{sec:catalog}

The cost law, the symmetry reduction, and the bond test together make it possible to enumerate the nanotubes of an entire materials space, complementing studies that map the full chirality space of a single compound such as \ce{MoS2}.\cite{Huang2026} We applied \ntbuilder{} to the relaxed monolayers of eight sources, namely C2DB,\cite{Haastrup2018, Gjerding2021} 2DMatPedia,\cite{Zhou2019} the planar carbon networks of Shi \textit{et al.},\cite{Shi2021} the Materials Cloud 2D database,\cite{Mounet2018, Campi2023} JARVIS-DFT 2D,\cite{Choudhary2017, Choudhary2020} the MXene data set of Ontiveros \textit{et al.},\cite{Ontiveros2023, Ontiveros2024, Ontiveros2025} MatHub-2d,\cite{Yao2023} and the 2D subset of Alexandria restricted to structures within 0.1~eV/atom of the convex hull,\cite{Wang2023, Cavignac2026} following the workflow of Figure~\ref{fig:catalog}.

The 54,496 structures read from the source databases were compared by a direct geometric overlay of their lattices and atomic positions, which merged duplicates across and within databases into 46,403 distinct 2D systems and 8,093 aliases. For each system, every irreducible index pair with a diameter between the thickness $t$ of the layer and $t + 30$~\AA\ was built in the smallest cell with a residual $\varepsilon$ of Eq.~\ref{eq:residual} of at most 0.5\% and at most 50,000 atoms, one tube per orbit of the group of Section~\ref{sec:symmetry} and one per inequivalent rolling sense. This procedure produced 14,609,330 index pairs and 20,334,372 nanotubes, whose distribution over the elements is shown in the lower part of Figure~\ref{fig:catalog}.

The bond test of Section~\ref{sec:curvature}, extended to both formed and broken bonds, classifies 57.1\% of the tubes as free of bond changes. Among the remaining ones, 19.9\% of all tubes only form new bonds, 6.7\% only break bonds of the flat layer, and 16.4\% do both. The fraction of unchanged tubes depends strongly on the thickness of the layer, as anticipated by Eq.~\ref{eq:dcrit}, reaching 99.7\% for the planar carbon networks and only 27.1\% for MXenes, whose multilayer slabs place their outer atoms far from the mid-plane. The catalog is accessible through a web page that filters tubes by elements, source database, diameter, number of atoms, and bond changes, delivers the requested structures in the export formats of \ntbuilder{}, and opens any catalog entry directly in the builder.

\section{Conclusions}
\label{sec:conclusions}

In summary, we have developed a general theory of the rolling construction for arbitrary 2D crystals and implemented it in \ntbuilder{}. The size of the exact unit cell of a tube follows from the reduced form of a single rational number built from the lattice metric, which explains why triangular and square layers always close with small cells while rectangular, centered rectangular, and oblique layers can require cells larger by many orders of magnitude, and the optimal approximate cells form a front of best rational approximations along which the compromise between size and periodicity is selected. The 17 plane groups induce seven distinct chirality maps once rolling is taken into account, and layers whose operations exchange their two faces, such as penta-graphene, tie the choice of chirality to the choice of rolling sense.

The effects of curvature are captured by a pair-level comparison with the flat layer, which identifies the bonds formed or broken by rolling and explains, for Janus \ce{MoSSe}, why the Se-inside sense changes bonds up to diameters twice as large as the S-inside sense. The same framework assembles multiwalled tubes whose walls share one axial period, periodic bundles, and strained or twisted fibers, and its application to 46,403 distinct 2D systems produced a public catalog of more than 20 million nanotubes. The catalog provides starting structures for first-principles and machine-learning studies of how the properties of 2D crystals change upon rolling, and the same construction can be extended to heterostructure tubes and nanoscrolls.\cite{Xiang2020, Braga2004, Perim2013}

\section*{Data and Software Availability}
The source code of \ntbuilder{} is available under the MIT License at \url{https://github.com/marcelo-lopes-pereira-junior/ntbuilder}. The web interface and the nanotube catalog are publicly accessible at \url{https://nanoeng.unb.br/ntbuilder/}.

\section*{Author Contributions}
M.L.P.J.: Conceptualization, Methodology, Software, Validation, Formal analysis, Investigation, Data Curation, Visualization, Writing - Original Draft, Writing - Review \& Editing, Funding acquisition.

\section*{Declaration of Competing Interests}
The author declares no known competing financial interests or personal relationships that could have appeared to influence the work reported in this paper.

\begin{acknowledgement}
M.L.P.J. acknowledges financial support from FAPDF (grant 00193-00001807/2023-16), CNPq (grants 444921/2024-9 and 308222/2025-3), and CAPES (grant 88887.005164/2024-00).
\end{acknowledgement}

\bibliography{bibliography}

\end{document}